\documentclass[showpacs,amssymb,preprint,preprintnumbers,nofootinbib,superscriptaddress]{revtex4}
\usepackage{amsmath}
\usepackage{graphicx}
\usepackage{latexsym}
\usepackage{amsfonts}
\usepackage{url,hyperref}
\usepackage{bm}
\usepackage{textcomp}
\usepackage{color}

\newcommand{\beq}{\begin{equation}}
\newcommand{\eeq}{\end{equation}}
\def\bea{\begin{eqnarray}}
\def\eea{\end{eqnarray}}

\begin{document}

\title{Stress-energy Tensor Correlators of a Quantum Field in Euclidean $R^N$ and $AdS^N$ spaces via the generalized zeta-function method}
\author{H. T. Cho}
\email[Email: ]{htcho@mail.tku.edu.tw}
\affiliation{Department of Physics, Tamkang University, Tamsui, New Taipei City, TAIWAN}
\author{B. L. Hu}
\email[Email: ]{blhu@umd.edu}
\affiliation{Maryland Center for Fundamental Physics, Department of Physics, University of Maryland, College Park, Maryland 20742-4111, USA}

\begin{abstract}
In this paper we calculate the vacuum expectation values of the stress-energy bitensor of a massive quantum scalar field with general coupling to N-dimensional Euclidean spaces and hyperbolic spaces which are Euclidean sections of the anti-de Sitter (AdS) spaces. These correlators, also known as the noise kernel, act as sources in the Einstein-Langevin equations of stochastic gravity \cite{HuVer08,HuVer03} which govern the induced metric fluctuations beyond the mean-field dynamics described by the semiclassical Einstein equations of semiclassical gravity. Because these spaces are  maximally symmetric the eigenmodes have analytic expressions which facilitate the computation of the  zeta-function \cite{DowCri76,Haw77}. Upon taking the second functional variation of the generalized  zeta function introduced in \cite{PH97} we obtain  the correlators of the stress tensor for these two classes of spacetimes.  Both the short and the large geodesic distance limits of the correlators are presented for dimensions up to 11. We mention current research problems in early universe cosmology, black hole physics and gravity-fluid duality where these results can be usefully applied to.
\end{abstract}

\pacs{04.62.+v, 05.40.-a, 11.10.Kk}
\date{\today}
\maketitle

%
%
\section{Introduction}\label{sec:intro}


In this paper we present a calculation of the  vacuum expectation value (vev) of the stress-energy bitensor of a massive quantum scalar field with general coupling to an N-dimensional Euclidean space and an N-dimensional hyperbolic space which is the Euclidean section of the anti-de Sitter space. Our results can be easily extended to an N-dimensional de Sitter space by analytical continuation.  There are at least three classes of physical problems of current interest which necessitate the knowledge of such quantities and motivated us to undertake this task. The theoretical framework which encompasses these research programs is known as (semiclassical) stochastic gravity \cite{HuVer03,HuVer08}.

Stochastic gravity (STG) is a theory established in the 90's  as a natural extension of semiclassical gravity theory (SCG) of the 80's \cite{scg}, for including the effects of fluctuations in the quantum matter field through the Einstein-Langevin equation (ELEq) \cite{ELE} which govern the behavior of the induced metric fluctuations \cite{HRV} \footnote{Note we make the distinction, and urge the community to do so as well, between ``fluctuations" which are classical stochastic variables or functions and ``perturbations" which obey classical deterministic equations. Thus in the quantum theory of structure formation the density contrast is related to (quantized) gravitational perturbations, not metric fluctuations.  Note also fluctuations are often conjured as quantum in nature as the difference from a classical background field when such a decomposition is carried out for a quantum field. Even so it remains a deterministic function obeying an operator equation, not a stochastic function. Metric fluctuations are related to spacetime foam and are often mentioned in connection with Planck scale physics, but in semiclassical stochastic gravity which is the only theory we are aware of which provides a quantitative description via the ELEq they remain as classical stochastic functions even though they are sourced by fluctuations of quantum matter fields. }. While SCG goes beyond quantum field theory in curved spacetime of the 70's \cite{BirDav} (QFTCST, viewed as the test-field approximation of SCG on a fixed background spacetime), in that the backreaction of the quantum matter field on the dynamics of the spacetime is incorporated through the expectation value of the stress energy tensor as the source of the semiclassical Einstein (SCE) equation,  stochastic gravity goes beyond SCG in that it includes also the backreaction of the fluctuations of the stress energy tensor, measured by the noise kernel (NK),  defined as the vacuum expectation value of the stress energy bitensor, which governs the dynamics of the induced metric fluctuations.

As applications of the stochastic gravity program and in particular the stress tensor correlator or the noise kernel to problems which contain important foundational physical issues, we mention   below two in cosmology, one  in  black hole physics and one in gravity-fluid duality.\\

\noindent {\bf Early Universe Cosmology}:

\noindent a) {\it Backreaction of quantum matter fields and their fluctuations in the early universe.}

Backreaction of quantum matter field is at the heart of both the SCG and STG program. Such effects have been studied in the late 70s and early 80s pertaining to possible removal of the cosmological singularity \cite{FHH}, the dissipation of anisotropy \cite{HarHu} and inhomogeneities \cite{CamVer94} and the decay of the cosmological constant \cite{TsaWoo,GarTan,Pol,Mottola}. The first level treatment of a backreaction problem relies on the calculation of the renormalized or regularized vev of the stress energy tensor which acts as the source in the SCE equations. The solutions of this equation begin the second level of a backreaction problem, where one begins with the calculation of the vev of the stress energy bitensor around these solutions as background spacetimes and then using them as sources  to seek the solutions of the ELEq  for the induced metric fluctuations.  An excellent example for this endeavor is the papers of Martin and Verdaguer \cite{MarVer00} for the metric fluctuations of the Minkowsky space. For consideration of issues like the decay of the cosmological constant in de Sitter-like universes the first step is undertaken by P\'erez-Nadalin, Roura and Verdaguer (PRV) \cite{PRVdS}  with the calculation of the correlators of stress energy for scalar fields in de Sitter spacetime. (Note, strictly speaking, for a backreaction problem,  the spacetime where such a quantity  need be calculated is not the de Sitter spacetime but the modified one from a solution of the SCE equations.)\\

\noindent  b) {\it Trace-anomaly driven inflation.} This scheme first proposed by Starobinsky \cite{StaInf} uses the trace anomaly as the source which drives the universe to an exponential expansion for a long enough duration to have interesting cosmological consequences \footnote{The equations derived by \cite{FHH} for the Robertson-Walker universe with  backreaction of the trace anomaly also contain such solutions, but these authors sought  boundary value solutions demanding that a radiation-dominated Friedmann universe exists at late times. If left as an initial value problem one would see unstable solutions not unlike what is obtained by Mottola et al \cite{Mottola}.}.  Its advantage over all later proposed inflation schemes rests on the fact that the trace anomaly is a necessary ingredient in semiclassical gravity arising from the renormalization or regularization of the vev of the energy momentum tensor of the matter fields.   Hawking, Hertog and Reall \cite{HHR} have considered this model in Euclidean space with the 4-sphere as the instanton solution. They used the AdS/CFT correspondence \cite{AdSCFT} to calculate the correlation functions for scalar and tensor metric perturbations and used them for structure formation considerations.

In   the quantum theory of structure formation the quantized linear gravitational perturbations (graviton) around a background spacetime are related to the inflaton perturbations around a background field configuration which in turn govern the density contrast (for a standard treatment, see, e.g., \cite{Feldman}). A viable quantum theory does not exist yet for treating quantized second order gravitational perturbations, but as explained by Roura and Verdaguer \cite{RouVerStrFor} one can reproduce these tensor perturbation correlators in the stochastic gravity program  by using their correspondence with the stochastic quantities, the equivalence for first order quantities are shown in \cite{CRV}.  For these second order stochastic quantities, one needs to calculate the noise kernel or the correlators of the stress energy tensor, which is the goal of this work.\\

\noindent {\bf Conformal Field Theory in maximally symmetric spaces of higher dimensions}

c) {\it Conformal field theories} (CFT) have deep and beautiful properties which capture the salient features of many physical systems \cite{CardyBook,CFTbooks}. In two dimension CFTs are characterized by a single quantity, the central charge $c$.  Parameters which are defined through correlation functions of the energy-momentum tensor, or its expectation value when the space on which the theory is defined has non-trivial topology or non-zero curvature, may serve to specify the theory independent of any particular formulation in terms of elementary fields. For example the central charge in the Virasoro algebra for two-dimensional CFTs may be defined through the trace of the energy-momentum tensor on a curved background. Such parameters should be well defined at any renormalisation group fixed point where the theory becomes conformal, such as in the case of free fields. Away from the critical points $c$ may be generalized to a function of the couplings, which monotonically decreases under renormalization group flow as the basic scale of
the theory is evolved to large distances and the couplings are attracted to any potential infrared fixed point. This is embodied in the Zamolodchikov c-theorem \cite{Zam86}.  That is why the short and long distance behavior of the correlators are of special physical significance.

It is of interest to inquire if such nice properties exist in higher dimensional CFTs, especially
the realistic four-dimensional spacetimes, as was first attempted by Cardy et al \cite{Cardy4D}.
A major undertaking by Osborn and Shore \cite{OsbSho00} provided a thorough analysis of one- and two-point functions of the energy-momentum tensor on homogeneous spaces of constant curvature. Their results for the two point functions provide a benchmark for all later derivations of such quantities in de Sitter space, as provided recently by \cite{PRVdS} and anti-de Sitter spaces, as in our present work. They have been used by Hawking et al \cite{HHR1} for the consideration of trace-anomaly driven inflation mentioned above.\\

\noindent {\bf  AdS black Hole Thermodynamics}

d) After the seminal work of Hawking \cite{Haw75} it is realized that with quantum effects black holes can evaporate. Black holes behave like thermal objects and can be in equilibrium with a thermal gas at the Hawking temperature. However, this equilibrium state is unstable to the absorption of the thermal gas by the black hole in an asymptotically flat spacetime \cite{Haw76}. In order to obtain an equilibrium state one has to put the black hole in an ad hoc cavity with some appropriate boundary condition \cite{Yor85}. The situation with black holes in an anti-de Sitter (AdS) spacetime is rather different as explored by Hawking and Page \cite{HP83}. Above the critical temperature there are two masses of the black hole which can be in equilibrium with the thermal gas. The larger mass black hole is actually stable. This phase transition is now known as the Hawking-Page transition.

The recent interest in this Hawking-Page transition is mainly due to the AdS/CFT correspondence \cite{Wit98}. This transition is conjectured to be related to the confinement phenomenon in QCD under this correspondence. Moreover, the quasinormal modes of the AdS black hole perturbations are supposed to give information of the thermal states of the conformal theory at the boundary \cite{HorHub}.

To study the Hawking-Page transition in a dynamical manner, the corresponding fluctuations of the quantum fields on the background spacetime must be taken into account. A systematic way to do so is by the theory of stochastic gravity \cite{HuVer08,HuVer03}. To the lowest order the fluctuations are represented by a noise kernel which is the vacuum expectation value of the correlation of the stress energy tensors. This is one of the reasons why we are interested in such a consideration here. For the AdS black hole case, one should deal with the stress tensor of a thermal gas in the Hartle-Hawking vacuum state. However, before taking such an endeavor we would like to get a grip on the problem by calculating the case with the scalar field in AdS spacetime. In our future work, we shall continue our investigation on the finite temperature case and then ultimately on the black hole spacetime.\\

\noindent {\bf Viscosity function in gravity fluid duality }

e) One form of AdS/CFT duality  has proven in the last decade to be of great interest to strong coupling gauge fluids as in RHIC experiments and cold atom physics, namely, the calculation of the viscosity function of a strongly interacting conformal gauge field (living on the boundary) by way of the corresponding quantity for a weakly coupled quantum field in an AdS spacetime (in the bulk). A viscosity to entropy density bound was obtained by Son et al \cite{Son} by means of the AdS/CFT correspondence. (See, e.g., their review \cite{SonRev}) Viscosity function obtained via the Kubo relation is restricted to linear response regimes, which assumes the system remains close to thermal equilibrium. Nonequilibrium expressions are more useful for a real time description of the fluid dynamics, but as explained by Son et al \cite{SonRev} its derivation is not so easy. We note that besides the direct method of quantum kinetic field theory \cite{CH88} via the closed-time-path or the Schwinger-Keldysh formalism (e.g., \cite{CHR,Jeon}) to obtain the transport functions (which is restricted to weak fields) there is another way to obtain the viscosity function, namely, via the influence functional approach \cite{IF} which can be applied to fully nonequilibrium systems. It is embedded in the dissipation kernel (the expectation value of the commutator of the stress tensor)
which is tethered with 
the noise kernel through the fluctuation-dissipation (FD) relation (see, e.g., \cite{HPZ,CRV,FDRscg}).  Earlier the polarization tensor for a finite-temperature quantum field far away from a black hole  (where the background metric is approximated by the Minkowski metric) was calculated by Campos and Hu \cite{CamHu} (see also Ref. [41]). Their results obtained from the Einstein-Langevin equations contain results  obtained by others via linear response theory as limiting case.  A similar calculation can be carried out for the AdS black hole where at large distances from the AdS-Schwarzschild horizon the spacetime approximates an AdS space. This requires an extension of the results reported here to finite temperatures which is underway.  Under fully nonequilibrium conditions there exist FD inequalities \cite{ChrisFDI} which contain useful information relating the  thermodynamic properties of the system of interest, such as dynamical susceptibility, to correlation functions in its environments.\\

\noindent {\bf Zeta function method in comparison to methods used earlier}

As mentioned above, the stress energy tensor correlators for scalar and fermions have been obtained before, notably, by Osborn and Shore
\cite{OsbSho00} for maximally symmetric spacetimes which include the de Sitter and anti-de Sitter spaces via conformal field theory techniques. However, they are only interested in the conformally coupled cases and have made extensive use of the traceless condition of the correlator. PRV have also calculated this quantity for quantum scalar fields in de Sitter spacetime. Their results can be analytically extended to AdS spaces and vice versa. On the other hand their results are for free minimally coupled scalar fields  while ours are for free fields with arbitrary couplings.

As this quantity is of basic importance for many reasons enumerated above it is useful to derive these expressions with a different method. The zeta function method we adopt here was introduced by Dowker and Critchley \cite{DowCri76} and Hawking \cite{Haw77}. It is one of the most elegant methods used for the regularization of the stress energy tensor for quantum fields in Riemannian (not pseudo-Riemannian)  spaces. One can use the zeta function to construct the regularized effective action and upon taking its first functional variation obtain the regularized stress energy tensor. (For a systematic exposition, see \cite{Elizalde}). This method is generalized by  Phillips and Hu \cite{PH97} to calculate the noise kernels of quantum scalar fields in $S^1 \times R^1$ spaces (useful for finite temperature theory and Casimir energy density considerations ) and for an Einstein universe, by way of the second order variation of the regularized effective action (see also \cite{CogEli02}). This method is used in our present calculation of the noise kernel for AdS spacetimes in N-dimensions.

In the next section, we shall describe the generalized zeta-function method of Phillips and Hu in some details as applied to a scalar field. Then in Section III, as a first application of the formalism, we calculate the expectation value of the stress-energy tensor in Euclideanized $AdS^N$ spaces. In Section IV, the correlators of the stress-energy tensors in both Euclidean $R^{N}$ and $AdS^{N}$ are calculated. The small and large geodesic distance expansions of the correlators in $AdS^{N}$ are given in Section V. The results are then compared with those in \cite{OsbSho00} and \cite{PRVdS}. Finally conclusions and discussions are presented in Section VI.



%
%

\section{Scalar field in spacetimes admitting an Euclidean section}\label{sec:scalar}

We consider a massive $m$ scalar field $\phi$ coupled to an $N$-dimensional Euclideanized space (with contravariant metric $g^{\mu\nu}(x)$, determinant $g$ and scalar curvature $R$) with coupling constant $\xi$ described by the action
\begin{eqnarray}
S[\phi]=\frac{1}{2}\int d^{N}x\sqrt{g(x)}\phi(x)H\phi(x),
\end{eqnarray}
where $H$ is the quadratic operator
\begin{eqnarray}
H=-\Box+m^{2}+\xi R,
\end{eqnarray}
and $R$ is the scalar curvature. The effective action defined by $W={\rm ln}{\cal Z}$ is related to
the generating functional ${\cal Z}$ by 
\begin{eqnarray}
{\cal Z}=\int{\cal D}\phi\ \! e^{-S[\phi]}.
\end{eqnarray}
The expectation value of the stress-energy tensor can be obtained by taking the functional derivative of the effective action
\begin{eqnarray}
\langle T_{\mu\nu}\rangle=-\frac{2}{\sqrt{g(x)}}\frac{\delta W}{\delta g^{\mu\nu}(x)}.
\end{eqnarray}
This formal expression is divergent at coincident limit and some procedure of regularization must be implemented. Here we adopt the procedure of $\zeta$-function regularization as shown by Dowker and Critchley \cite{DowCri76} and Hawking \cite{Haw77}. As shown by Phillips and Hu \cite{PH97} the fluctuation of the stress-energy tensor is obtained by taking another derivatives of the regularized effective action $W$,
\begin{eqnarray}
\Delta T^{2}_{\mu\nu\alpha'\beta'}(x,x')&\equiv&\langle T_{\mu\nu}(x)T_{\alpha'\beta'}(x')\rangle-\langle T_{\mu\nu}(x)\rangle\langle T_{\alpha'\beta'}(x')\rangle\nonumber\\
&=&\frac{4}{\sqrt{g(x)g(x')}}\frac{\delta^{2}W}{\delta g^{\mu\nu}(x)\delta g^{\alpha'\beta'}(x')}
\end{eqnarray}
Note that although the regularized expectation value of the stress-energy tensor is dependent on one spacetime point, the fluctuation of the stress energy is a  bitensor defined at two separate spacetime points through the two functional derivatives taken with respect to these two separate spacetime points.

Conventionally one define the $\zeta$-function of an operator $H$,
\begin{eqnarray}
\zeta_{H}(s)=\sum_{n}\left(\frac{\mu}{\lambda_{n}}\right)^{s}={\rm Tr}\left(\frac{\mu}{H}\right)^{s}
\end{eqnarray}
where $\lambda_{n}$ is the eigenvalues of $H$ and $\mu$ represents the renormalization scale.
The $\zeta$-function regularized effective action of the operator $H$ is
\begin{eqnarray}
W_{R}=\left.\frac{1}{2}\frac{d\zeta}{ds}\right|_{s\rightarrow 0}.
\end{eqnarray}
Using the proper-time method \cite{DowCri76}  one can write the $\zeta$-function as
\begin{eqnarray}
\zeta_{H}(s)&=&\frac{\mu^{s}}{\Gamma(s)}\int_{0}^{\infty}dt\ \! t^{s-1}{\rm Tr}e^{-tH}\\
W_{R}&=&\frac{1}{2}\frac{d}{ds}\left[\frac{\mu^{s}}{\Gamma(s)}\int_{0}^{\infty}dt\ \! t^{s-1}{\rm Tr}e^{-tH}\right]_{s\rightarrow 0}
\end{eqnarray}
Taking the first variation of the $\zeta$-function,
\begin{eqnarray}
\delta\zeta_{H}&=&-\frac{\mu^{s}}{\Gamma(s)}\int_{0}^{\infty}dt\ \! t^{s}{\rm Tr}\left(\delta He^{-tH}\right)\nonumber\\
&=&-\frac{\mu^{s}}{\Gamma(s)}\int_{0}^{\infty}dt\ \! t^{s}\sum_{n}e^{-t\lambda_{n}}\langle n\left|\delta H\right|n\rangle
\end{eqnarray}
we obtain the regularized expectation value of the stress-energy tensor which is given by
\begin{eqnarray}
\langle T_{\mu\nu}(x)\rangle&=&\frac{1}{2}\frac{d}{ds}\left[-\frac{\mu^{s}}{\Gamma(s)}\int_{0}^{\infty}dt\ \! t^{s}\sum_{n}e^{-t\lambda_{n}}\left(-\frac{2}{\sqrt{g(x)}}\left\langle n\right|\frac{\delta H}{\delta g^{\mu\nu}(x)}\left|n\right\rangle\right)\right]_{s\rightarrow 0}\nonumber\\
&=&-\frac{1}{2}\frac{d}{ds}\left\{\frac{\mu^{s}}{\Gamma(s)}\int_{0}^{\infty}dt\ \! t^{s}\sum_{n}e^{-t\lambda_{n}}\, T_{\mu\nu}\left[\phi_{n}(x),\phi_{n}^{*}(x)\right]\right\}_{s\rightarrow 0}\label{renstress}
\end{eqnarray}
where
\begin{eqnarray}
T_{\mu\nu}\left[\phi_{n}(x),\phi_{n'}^{*}(x)\right]&\equiv&-\frac{2}{\sqrt{g(x)}}\left\langle n'\right|\frac{\delta H}{\delta g^{\mu\nu}(x)}\left|n\right\rangle\nonumber\\
&=&-\frac{2}{\sqrt{g(x)}}\int d^{N}x'\sqrt{g(x')}\phi_{n'}^{*}(x')\left[\frac{\delta H}{\delta g^{\mu\nu}(x)}\phi_{n}(x')\right]
\end{eqnarray}
and $\phi_{n}(x)$ is a eigenfunction of the operator $H$, namely,
\begin{equation}
H \phi_n = \lambda_n \phi_n
\end{equation}
where $ \lambda_n $ are the eigenvalues corresponding to the eigenfunctions $\phi_n$.
We then have
\begin{eqnarray}
T_{\mu\nu}\left[\phi_{n}(x),\phi_{n'}^{*}(x)\right]&=&
-\left(\partial_{\mu}\phi_{n'}^{*}\partial_{\nu}\phi_{n}+\partial_{\nu}\phi_{n'}^{*}\partial_{\mu}\phi_{n}\right)
+g_{\mu\nu}\left(g^{\alpha\beta}\partial_{\alpha}\phi_{n'}^{*}\partial_{\beta}\phi_{n}+\phi_{n'}^{*}\Box\phi_{n}\right)\nonumber\\
&&\ \ \ -2\xi\left[g_{\mu\nu}\Box(\phi_{n'}^{*}\phi_{n})-\nabla_{\mu}\nabla_{\nu}(\phi_{n'}^{*}\phi_{n})+R_{\mu\nu}\phi_{n'}^{*}\phi_{n}\right]
\end{eqnarray}

In the approach of Phillips and Hu \cite{PH97}, through the use of the Schwinger method \cite{Sch51}, the second variation of the $\zeta$-function can be written as
\begin{eqnarray}
\delta_{2}\delta_{1}\zeta_{H}&=&\frac{\mu^{s}}{2\Gamma(s)}\int_{0}^{\infty}du\int_{0}^{\infty}dv(u+v)^{s}(uv)^{\nu}\nonumber\\
&&\ \ \ \ \ \left\{{\rm Tr}\left[(\delta_{1}H) e^{-uH}(\delta_{2}H) e^{-vH}\right]+{\rm Tr}\left[(\delta_{2}H) e^{-uH}(\delta_{1}H) e^{-vH}\right]\right\}
\end{eqnarray}
and
\begin{eqnarray}
\Delta T_{\mu\nu\alpha'\beta'}^{2}(x,x')&=&\frac{1}{2}\frac{d}{ds}\left\{\frac{\mu^{s}}{\Gamma(s)}
\int_{0}^{\infty}du\int_{0}^{\infty}dv(u+v)^{s}(uv)^{\nu}\sum_{n,n'}e^{-u\lambda_{n}-v\lambda_{n'}}\right.\nonumber\\
&&\ \ \ \ \ \ \left. T_{\mu\nu}\left[\phi_{n}(x),\phi_{n'}^{*}(x)\right]T_{\alpha'\beta'}\left[\phi_{n'}(x'),\phi_{n}^{*}(x')\right]\right.\!{\Bigg \}}_{s,\nu\rightarrow 0}.\label{T2exp}
\end{eqnarray}
Note that in this Phillips-Hu prescription \cite{PH97}, an additional regularization factor $(uv)^{\nu}$ has been introduced. This is because the authors were interested in the fluctuations of the stress-energy tensor, that is, in the coincident limit of $\Delta T_{\mu\nu\alpha'\beta'}^{2}(x,x')$ where under this limit further divergences occur which call for an additional regularization factor. (See also \cite{CogEli02}). However,  our present purpose is focused on getting the correlators with two points separated, i.e., in the non-coincident case. Hence, apart from the fact that the expression in Eq.~(\ref{T2exp}) is more symmetric with this factor, the keeping of this factor above is actually a matter of convenience. Here we can first take the $s\rightarrow 0$ limit without spoiling the regularization and the expression in Eq.~(\ref{T2exp}) will become
\begin{eqnarray}
\Delta T_{\mu\nu\alpha'\beta'}^{2}(x,x')&=&\frac{1}{2}
\int_{0}^{\infty}du\int_{0}^{\infty}dv (uv)^{\nu}\sum_{n,n'}e^{-u\lambda_{n}-v\lambda_{n'}}\nonumber\\
&&\ \ \ \ \ \ T_{\mu\nu}\left[\phi_{n}(x),\phi_{n'}^{*}(x)\right]T_{\alpha'\beta'}
\left[\phi_{n'}(x'),\phi_{n}^{*}(x')\right]\!{\bigg |}_{\nu\rightarrow 0}.\label{T2finalexp}
\end{eqnarray}
We shall see from the following calculations that with this expression the integrations over $u$ and $v$ effectively separate. The calculations are therefore simplified considerably.


%
%

\section{Expectation value of the stress-energy tensor in AdS spaces}\label{sec:Expect}

In this section we would like to apply the $\zeta$-function formalism in Section~\ref{sec:scalar} to the case of scalar fields in N-dim AdS spaces. The Euclideanized N-dim AdS space is the hyperbolic space $H^{N}$ with the metric
\begin{eqnarray}
ds^{2}=d\sigma^{2}+a^{2}\sinh^{2}\left(\frac{\sigma}{a}\right)d\Omega_{N-1}^{2}
\end{eqnarray}
where $\sigma$ is the geodesic distance,
$a$ the radius of the $AdS^N$ space, and $d\Omega_{N-1}^{2}$ the metric for the $(N-1)$-sphere. The eigenfunctions $\phi_{\kappa lm}$ obey the equations,
\begin{eqnarray}
\left(-\Box-\frac{\rho_{N}}{a^{2}}\right)\phi_{\kappa lm}=\left(\frac{\kappa^{2}}{a^{2}}\right)\phi_{\kappa lm}
\end{eqnarray}
where $\rho_{N}=(N-1)/2$, and are given by
\begin{eqnarray}
\phi_{\kappa lm}&=&c_{l}(\kappa)\left(\sinh\frac{\sigma}{a}\right)^{1-\frac{N}{2}}
P_{-\frac{1}{2}+i\kappa}^{1-l-\frac{N}{2}}\left(\cosh\frac{\sigma}{a}\right)Y_{lm}(\Omega)\nonumber\\
&\equiv&f_{\kappa l}(\sigma)Y_{lm}(\Omega)
\end{eqnarray}
where $P_{\nu}^{\mu}(x)$ is the associated Legendre function and $Y_{lm}(\Omega)$ are the hyperspherical harmonics. The normalization constant is given by
\begin{eqnarray}
c_{l}(\kappa)=\frac{|\Gamma(i\kappa+\rho_{N})|}{|\Gamma(i\kappa)|}
\end{eqnarray}
Later we shall need the normalization constant for $l=0$. For odd $N$,
\begin{eqnarray}
|c_{0}(\kappa)|^{2}=\frac{1}{a^{N}}\prod_{j=0}^{\rho_{N}}\left(\kappa^{2}+j^{2}\right)
=\frac{1}{a^{N}}\sum_{n=1}^{\rho_{N}}c_{2n}\kappa^{2n}\label{c0odd}
\end{eqnarray}
and for even $N$,
\begin{eqnarray}
|c_{0}(\kappa)|^{2}=\frac{1}{a^{N}}(\kappa\tanh\pi\kappa)\prod_{j=\frac{1}{2}}^{\rho_{N}}\left(\kappa^{2}+j^{2}\right)
=\frac{1}{a^{N}}\tanh\pi\kappa\sum_{n=0}^{\rho_{N}-\frac{1}{2}}c_{2n+1}\kappa^{2n+1}\label{c0even}
\end{eqnarray}
In both cases we have turned it into a finite sum.

Now we are ready to evaluate the regularized expectation value of the stress-energy tensor in AdS spaces. From Eq.~(\ref{renstress}),
\begin{eqnarray}
&&\langle T_{\mu\nu}(x)\rangle\nonumber\\
&=&-\frac{1}{2}\frac{d}{ds}\left\{\frac{\mu^{s}}{\Gamma(s)}\int_{0}^{\infty}dt\ \! t^{s}
\int_{0}^{\infty}d\kappa\sum_{lm}e^{-\frac{t}{a^{2}}\left(\kappa^{2}+\rho_{N}^{2}+m^{2}a^{2}-\xi N(N-1)\right)}\, T_{\mu\nu}\left[\phi_{\kappa lm}(x),\phi_{\kappa lm}^{*}(x)\right]\right\}_{s\rightarrow 0}\nonumber\\
\end{eqnarray}
since $\kappa$ is a continuous variable and $l$ and $m$ are discrete.
$H^{N}$ is a homogeneous space so $\langle T_{\mu\nu}(x)\rangle=Fg_{\mu\nu}(x)$ where $F$ is a constant. It is obvious that
\begin{eqnarray}
F&=&\left.\langle T_{\sigma\sigma}(x)\rangle\right|_{x\rightarrow 0}\nonumber\\
&=&-\frac{1}{2}\frac{d}{ds}\left\{\frac{\mu^{s}}{\Gamma(s)}\int_{0}^{\infty}dt\ \! t^{s}
\int_{0}^{\infty}d\kappa\sum_{lm}e^{-\frac{t}{a^{2}}\left(\kappa^{2}+a^{2}b\right)}\, T_{\sigma\sigma}\left[\phi_{\kappa lm}(x),\phi_{\kappa lm}^{*}(x)\right]\right\}_{x,s\rightarrow 0}\nonumber\\ \label{F}
\end{eqnarray}
where $a^{2}b=\rho^{2}_{N}+m^{2}a^{2}-\xi N(N-1)$. Also, we have taken the limit $x\rightarrow 0$ to simplify the evaluation.
$T_{\sigma\sigma}\left[\phi_{\kappa lm}(x),\phi_{\kappa lm}^{*}(x)\right]$ can be further simplified as
\begin{eqnarray}
&&T_{\sigma\sigma}\left[\phi_{\kappa lm}(x),\phi_{\kappa lm}^{*}(x)\right]\nonumber\\
&=&-2\partial_{\sigma}\phi_{\kappa lm}^{*}\partial_{\sigma}\phi_{\kappa lm}
+\left[-2\left(\xi-\frac{1}{4}\right)\Box+2\xi\nabla_{\sigma}\nabla_{\sigma}+2\xi\left(\frac{N-1}{a^{2}}\right)\right]
\left|\phi_{\kappa lm}\right|^{2}\label{tss}
\end{eqnarray}

Making use of the addition theorem of the hyperspherical harmonics,
\begin{eqnarray}
\sum_{m}Y^{*}_{lm}(\Omega)Y_{lm}(\Omega')=\frac{(2l+N-2)\Gamma\left(\frac{N-2}{2}\right)}{4\pi^{\frac{N}{2}}}
C_{l}^{\frac{N-2}{2}}\left(\Omega\cdot\Omega'\right)\label{addtheorem}
\end{eqnarray}
where $C_{l}^{n}$ is the Gegenbauer polynomial, we have
\begin{eqnarray}
\sum_{m}\left|\phi_{\kappa lm}\right|^{2}=\frac{(2l+N-2)\Gamma(l+N-2)}{2^{N-1}\pi^{\frac{N-1}{2}}\Gamma\left(\frac{N-1}{2}\right)\Gamma(l+1)}
|c_{l}(\kappa)|^{2}\left[\left(\sinh\frac{\sigma}{a}\right)^{1-\frac{N}{2}}
P_{-\frac{1}{2}+i\kappa}^{1-l-\frac{N}{2}}\left(\cosh\frac{\sigma}{a}\right)\right]^{2}
\end{eqnarray}
Moreover, from the short distance expansion of $\phi_{\kappa lm}$,
\begin{eqnarray}
&&\left(\sinh\frac{\sigma}{a}\right)^{1-\frac{N}{2}}
P_{-\frac{1}{2}+i\kappa}^{1-l-\frac{N}{2}}\left(\cosh\frac{\sigma}{a}\right)\nonumber\\
&=&\frac{2^{1-l-\frac{N}{2}}}{\Gamma\left(l+\frac{N}{2}\right)}
\left(\frac{\sigma}{a}\right)^{l}\left\{1-\left[\frac{l}{6}+\frac{\kappa^{2}+(\rho_{N}+l)^{2}}{2(2l+N)}\right]\left(\frac{\sigma}{a}\right)^{2}
+\cdots\right\}
\end{eqnarray}
only the $l=0$ and the $l=1$ terms will contribute to the $l$ sum in the limit $x$ or $\sigma\rightarrow 0$. In particular,
\begin{eqnarray}
\left(\sum_{m}|\phi_{\kappa 0m}|^{2}\right)_{\sigma\rightarrow 0}&=&\frac{|c_{0}(\kappa)|^{2}}{2^{N-1}\pi^{\frac{N}{2}}\Gamma\left(\frac{N}{2}\right)}\\
\left(\sum_{m}\partial_{\sigma}\phi_{\kappa 1m}^{*}\partial_{\sigma}\phi_{\kappa 1m}\right)_{\sigma\rightarrow 0}
&=&\frac{(\kappa^{2}+\rho_{N}^{2})|c_{0}(\kappa)|^{2}}{2^{N-1}\pi^{\frac{N}{2}}N\Gamma\left(\frac{N}{2}\right)a^{2}}\\
\left(\Box\sum_{m}|\phi_{\kappa 0m}|^{2}\right)_{\sigma\rightarrow 0}&=&-\frac{(\kappa^{2}+\rho_{N}^{2})|c_{0}(\kappa)|^{2}}
{2^{N-2}\pi^{\frac{N}{2}}N\Gamma\left(\frac{N}{2}\right)a^{2}}\\
\left(\Box\sum_{m}|\phi_{\kappa 1m}|^{2}\right)_{\sigma\rightarrow 0}&=&\frac{(\kappa^{2}+\rho_{N}^{2})|c_{0}(\kappa)|^{2}}
{2^{N-2}\pi^{\frac{N}{2}}N\Gamma\left(\frac{N}{2}\right)a^{2}}
\end{eqnarray}
Hence,
\begin{eqnarray}
\left(\Box\sum_{lm}|\phi_{\kappa lm}|^{2}\right)_{x\rightarrow 0}
=\left(\nabla_{\sigma}\nabla_{\sigma}\sum_{lm}|\phi_{\kappa lm}|^{2}\right)_{x\rightarrow 0}=0
\end{eqnarray}
and we have
\begin{eqnarray}
\left(\sum_{lm}T_{\sigma\sigma}[\phi_{\kappa lm}(x),\phi_{\kappa lm}^{*}(x)]\right)_{x\rightarrow 0}
=-\frac{|c_{0}(\kappa)|^{2}}{2^{N-2}\pi^{\frac{N}{2}}N\Gamma\left(\frac{N}{2}\right)a^{2}}[\kappa^{2}+\rho_{N}^{2}-\xi N(N-1)]
\end{eqnarray}
The constant $F$ in Eq.~(\ref{F}) becomes
\begin{eqnarray}
F=\frac{1}{2^{N-1}\pi^{\frac{N}{2}}N\Gamma\left(\frac{N}{2}\right)}\frac{d}{ds}\left\{\frac{\mu^{s}a^{2s}}{\Gamma(s)}\int_{0}^{\infty}dt\ \! t^{s}
\int_{0}^{\infty}d\kappa\ \! e^{-t(\kappa^{2}+a^{2}b)}|c_{0}(\kappa)|^{2}(\kappa^{2}+a^{2}b-m^{2}a^{2})\right\}_{s\rightarrow 0}
\end{eqnarray}
Using the finite sum representation for $|c_{0}(\kappa)|^{2}$ in the case of odd dimensions in Eq.~(\ref{c0odd}), we have
\begin{eqnarray}
F=\frac{m^{2}a^{2-N}}{2^{N}\pi^{\frac{N}{2}-1}N\Gamma\left(\frac{N}{2}\right)}\sum_{n=1}^{\rho_{N}}(-1)^{n+1}c_{2n}(a^{2}b)^{n-\frac{1}{2}}
\end{eqnarray}
For even dimensions, $|c_{0}(\kappa)|^{2}$ involves the term $\tanh\pi\kappa=1-2/(e^{2\pi\kappa}+1)$, and one has
\begin{eqnarray}
F&=&\frac{a^{-N}}{2^{N-1}\pi^{\frac{N}{2}}N\Gamma\left(\frac{N}{2}\right)}\sum_{n=0}^{\rho_{N}-\frac{1}{2}}c_{2n+1}
\left[\frac{(-1)^{n+1}}{2(n+1)}(a^{2}b)^{n+1}\right.\nonumber\\
&&\ \ \left.+\frac{(-1)^{n+1}}{2}(a^{2}b)^{n}m^{2}a^{2}\left(d_{n}-{\rm ln}\frac{b}{\mu}\right)+2m^{2}a^{2}H_{n}(1;\sqrt{a}b)-2H_{n}(0)\right]
\end{eqnarray}
where
\begin{eqnarray}
d_{0}=0,\ \ \ \ \ d_{n}=\sum_{k=1}^{n}\frac{1}{k}\ \ \ (n\geq 1),
\end{eqnarray}
and the function $H_{n}(s;\mu)$ is defined by the integral
\begin{eqnarray}
H_{n}(s;\mu)=\int_{0}^{\infty}\frac{\kappa^{2n+1}d\kappa}{(e^{2\pi\kappa}+1)(\kappa^{2}+\mu^{2})^{s}}
\end{eqnarray}
Our result is the same as in \cite{Cal99} in which a different coordinate system has been used.

%
%

\section{Correlations of the stress-energy tensor}\label{sec:fluctuations}

In a maximally symmetric space like the Euclidean space $R^N$ or the hyperbolic space $H^{N}$, any bitensor can be expressed in terms of a set of basic bitensors \cite{AllJac86}. The first basic bitensor is the bi-scalar function $\tau(x,x')$
which is the geodesic distance between $x$ and $x'$. Using the covariant derivative one could define
\begin{equation}
n_{\mu}=\nabla_{\mu}\tau(x,x')\ \ \ ;\ \ \ n_{\alpha'}=\nabla_{\alpha'}\tau(x,x')
\end{equation}
where $n_{\mu}(x,x')$ is a vector at $x$ and a scalar at $x'$, while $n_{\alpha'}(x,x')$ is a scalar at $x$ and a vector at $x'$. Next, we have the parallel propagator ${g_{\mu}}^{\alpha'}(x,x')$ which parallel transports any vector $v^{\mu}$ from $x$ to $x'$. The transported vector is $v^{\mu}{g_{\mu}}^{\alpha'}(x,x')$. It is easy to see that $n_{\mu}(x,x')=-{g_{\mu}}^{\alpha'}n_{\alpha'}(x,x')$. $\tau(x,x')$, $n_{\mu}(x,x')$, $n_{\alpha'}(x,x')$ and ${g_{\mu}}^{\alpha'}(x,x')$ constitute this set of basic bitensors. All other bitensors can be expressed in terms of them with coefficients depending only on the geodesic distance between the two points. For example,
\begin{eqnarray}
\nabla^{\mu}n_{\nu}&=&A(\tau)\left({g^{\mu}}_{\nu}-n^{\mu}n_{\nu}\right)\label{der1}\\
\nabla^{\mu}n_{\alpha'}&=&B(\tau)\left({g^{\mu}}_{\alpha'}+n^{\mu}n_{\alpha'}\right)\label{der2}\\
\nabla^{\mu}g_{\nu\alpha'}&=&-(A(\tau)+B(\tau))\left({g^{\mu}}_{\nu}n_{\alpha'}+{g^{\mu}}_{\alpha'}n_{\nu}\right)\label{der3}
\end{eqnarray}
where $A(\tau)$ and $B(\tau)$ are function of $\tau$ only. For the Euclidean $R^{N}$ spaces, $A=1/\tau$ and $B=-1/\tau$. For hyperbolic $H^{N}$ spaces, $A=\coth(\tau/a)/a$ and $B=-{\rm csch}(\tau/a)/a$.

Since $\Delta T_{\mu\nu\alpha'\beta'}^{2}(x,x')$ is a symmetric bitensor, one could express it in terms of these basic bitensors. Taking the symmetries of the indices into account, we have
\begin{eqnarray}
\Delta T_{\mu\nu\alpha'\beta'}^{2}(x,x')&=&C_{1}(\tau)n_{\mu}n_{\nu}n_{\alpha'}n_{\beta'}
+C_{2}(\tau)\left(n_{\mu}n_{\nu}g_{\alpha'\beta'}+g_{\mu\nu}n_{\alpha'}n_{\beta'}\right)\nonumber\\ &&\ \ +C_{3}(\tau)\left(n_{\mu}g_{\nu\alpha'}n_{\beta'}+n_{\nu}g_{\mu\alpha'}n_{\beta'}+n_{\mu}g_{\nu\beta'}n_{\alpha'} +n_{\nu}g_{\mu\beta'}n_{\alpha'}\right)\nonumber\\
&&\ \ +C_{4}(\tau)\left(g_{\mu\alpha'}g_{\nu\beta'}+g_{\mu\beta'}g_{\nu\alpha'}\right)+C_{5}(\tau)g_{\mu\nu}g_{\alpha'\beta'}\label{T2bitensor}
\end{eqnarray}
Using the derivatives in Eqs.~(\ref{der1}) to (\ref{der3}), the conservation conditions $\nabla^{\mu}\Delta T_{\mu\nu\alpha'\beta'}^{2}(x,x')=0$ can be expressed as three equations on the coefficients,
\begin{eqnarray}
&&\frac{dC_{1}}{d\tau}+\frac{dC_{2}}{d\tau}-2\frac{dC_{3}}{d\tau}+(N-1)AC_{1}+2BC_{2}-2\left((N-2)A+NB\right)C_{3}=0\nonumber\\
&&\frac{dC_{2}}{d\tau}+\frac{dC_{5}}{d\tau}+(N-1)AC_{2}+2BC_{3}-2(A+B)C_{4}=0\nonumber\\
&&\frac{dC_{3}}{d\tau}-\frac{dC_{4}}{d\tau}+BC_{2}+NAC_{3}-N(A+B)C_{4}=0\label{conservation}
\end{eqnarray}
Moreover, the traceless condition $g^{\mu\nu}\Delta T_{\mu\nu\alpha'\beta'}^{2}(x,x')=0$ can be written as
\begin{eqnarray}
C_{1}+NC_{2}-4C_{3}&=&0\nonumber\\
C_{2}+2C_{4}+NC_{5}&=&0\label{conformal}
\end{eqnarray}

\subsection{Stress-energy tensor correlators in Euclidean flat space $R^N$}

We begin with a calculation of the correlator $\Delta T_{\mu\nu\alpha\beta}^{2}(x,x')$ of the stress tensor in the Euclidean $R^N$  space which can be presented in closed form. The corresponding eigenfunction is just
\begin{equation}
\phi_{\vec{k}}(\vec{x})=\frac{1}{(2\pi)^{N/2}}e^{i\vec{k}\cdot\vec{x}}
\end{equation}
In this case the correlator in Eq.~(\ref{T2finalexp}) can be simplified to
\begin{eqnarray}
&&\Delta T_{\mu\nu\alpha'\beta'}^{2}(x,x')\nonumber\\
&=&\frac{1}{2}\int_{0}^{\infty}du u^{\nu}\int_{0}^{\infty}dv v^{\nu}
\int\frac{d^{N}k}{(2\pi)^{N}}\int\frac{d^{N}k'}{(2\pi)^{N}}e^{-u(\vec{k}^{2}+m^{2})}e^{-v(\vec{k}'^{2}+m^{2})}
e^{i\vec{k}\cdot(\vec{x}-\vec{x}')}e^{-i\vec{k}'\cdot(\vec{x}-\vec{x}')}\nonumber\\
&&\ \ \left\{-(k_{\mu}k'_{\nu}+k'_{\mu}k_{\nu})-\delta_{\mu\nu}\vec{k}\cdot(\vec{k}-\vec{k}')+2\xi\left[\delta_{\mu\nu}(\vec{k}-\vec{k}')^{2}
-(k-k')_{\mu}(k-k')_{\nu}\right]\right\}\nonumber\\
&&\ \ \left\{-(k_{\alpha'}k'_{\beta'}+k'_{\alpha'}k_{\beta'})+\delta_{\alpha'\beta'}\vec{k}'\cdot(\vec{k}-\vec{k}')
+2\xi\left[\delta_{\alpha'\beta'}(\vec{k}-\vec{k}')^{2}-(k-k')_{\alpha'}(k-k')_{\beta'}\right]\right\}\nonumber\\
\end{eqnarray}

After doing the integrations one can indeed expressed $\Delta T_{\mu\nu\alpha'\beta'}^{2}(x,x')$ in terms of the set of basic bitensors as in Eq.~(\ref{T2bitensor}). In this flat case, $\tau=|x-x'|$ , $n_{\mu}=(x-x')_{\mu}/|x-x'|$, $n_{\alpha'}=(x'-x)_{\alpha'}/|x-x'|$, and $g_{\mu\alpha'}=\delta_{\mu\alpha'}$. The coefficients $C_{1}$ to $C_{5}$ can be expressed in terms of the modified Bessel functions $K_{\nu}(z)$,
\begin{eqnarray}
C_{1}&=&\frac{2}{(2\pi)^{N}}\left(\frac{m^{N-1}}{\tau^{N+1}}\right)
\bigg\{m\tau\left[2(N+4)(N+2)\xi^{2}+m^{2}\tau^{2}\left(1-4\xi+8\xi^{2}\right)\right]
\left(K_{\frac{N}{2}-1}\left(m\tau\right)\right)^{2}\nonumber\\
&&\ \ \ +2\big[N(N+2)(N+4)\xi^{2}\nonumber\\
&&\ \ \ \ \ \ \ \ \ \ \ \ \ \ \ +m^{2}\tau^{2}\left[N-2(3N+2)\xi+12(N+1)\xi^{2}\right]\big]
K_{\frac{N}{2}}\left(m\tau\right)K_{\frac{N}{2}-1}\left(m\tau\right)\nonumber\\
&&\ \ \ m\tau\left[N\left(N-8(N+1)\xi+2(7N+8)\xi^{2}\right)-4m^{2}\tau^{2}\xi(1-2\xi)\right]
\left(K_{\frac{N}{2}}\left(m\tau\right)\right)^{2}\bigg\}\\
C_{2}&=&-\frac{1}{(2\pi)^{N}}\left(\frac{m^{N-1}}{\tau^{N+1}}\right)
\bigg\{m\tau\left[4(N+2)\xi^{2}+m^{2}\tau^{2}(1-4\xi)^{2}\right]\left(K_{\frac{N}{2}-1}\left(m\tau\right)\right)^{2}
\nonumber\\
&&\ \ \ +2\left[2N(N+2)\xi^{2}-m^{2}\tau^{2}(1-4\xi)(1-N+4N\xi)\right]
K_{\frac{N}{2}}\left(m\tau\right)K_{\frac{N}{2}-1}\left(m\tau\right)\nonumber\\
&&\ \ \ +m\tau\left[N\left((N-2)-8(N-1)\xi+4(4N-1)\xi^{2}\right)+m^{2}\tau^{2}(1-4\xi)^{2}\right]
\left(K_{\frac{N}{2}}\left(m\tau\right)\right)^{2}\bigg\}\nonumber\\ \\
C_{3}&=&\frac{1}{(2\pi)^{N}}\left(\frac{m^{N-1}}{\tau^{N+1}}\right)
\bigg\{m\tau\left[4(N+2)\xi^{2}\right]\left(K_{\frac{N}{2}-1}\left(m\tau\right)\right)^{2}\nonumber\\
&&\ \ \ \ \ \ \ \ \ \ +\left[4N(N+2)\xi^{2}+m^{2}\tau^{2}(1-4\xi)^{2}\right]
K_{\frac{N}{2}}\left(m\tau\right)K_{\frac{N}{2}-1}\left(m\tau\right)\nonumber\\
&&\ \ \ \ \ \ \ \ \ \ +m\tau N(1-8\xi+12\xi^{2})
\left(K_{\frac{N}{2}}\left(m\tau\right)\right)^{2}\bigg\}\\
C_{4}&=&\frac{1}{(2\pi)^{N}}\left(\frac{m^{N-1}}{\tau^{N+1}}\right)
\bigg\{m\tau(4\xi^{2})\left(K_{\frac{N}{2}-1}\left(m\tau\right)\right)^{2}+4N\xi^{2}
K_{\frac{N}{2}}\left(m\tau\right)K_{\frac{N}{2}-1}\left(m\tau\right)\nonumber\\
&&\ \ \ \ \ \ \ \ \ \ \ \ \ \ \ +m\tau (1-2\xi)^{2}
\left(K_{\frac{N}{2}}\left(m\tau\right)\right)^{2}\bigg\}
\end{eqnarray}
\begin{eqnarray}
C_{5}&=&\frac{1}{2(2\pi)^{N}}\left(\frac{m^{N-1}}{\tau^{N+1}}\right)
\bigg\{2m\tau\left[4\xi^{2}+m^{2}\tau^{2}(1-4\xi)^{2}\right]\left(K_{\frac{N}{2}-1}\left(m\tau\right)\right)^{2}\nonumber\\
&&\ \ \ +2\left[4N\xi^{2}+m^{2}\tau^{2}(1-4\xi)((N-1)-4(N+1)\xi)\right]
K_{\frac{N}{2}}\left(m\tau\right)K_{\frac{N}{2}-1}\left(m\tau\right)\nonumber\\
&&\ \ \ +m\tau\left[\left((N^{2}-N-4)-8(N^{2}-3)\xi+8(2N^{2}+2N-3)\xi^{2}\right)\right.\nonumber\\
&&\ \ \ \ \ \ \ \ \ \ \left.+2m^{2}\tau^{2}(1-4\xi)^{2}\right]
\left(K_{\frac{N}{2}}\left(m\tau\right)\right)^{2}\bigg\}
\end{eqnarray}

From the properties of $K_{\nu}(z)$, one can show that these coefficients satisfy the conservation conditions as in Eqs.~(\ref{conservation}). To consider the traceless conditions we first list the coefficients in the case of massless scalar. However, for $N=2$ this limit is singular, so we present the result for the conformal scalar with $m=\xi=0$. In this case,
\begin{equation}
C_{1}=\frac{2}{\pi^{2}\tau^{4}}\ \ \ ;\ \ \ C_{2}=0\ \ \ ;\ \ \ C_{3}=\frac{1}{2\pi^{2}\tau^{4}}\ \ \ ;\ \ \ C_{4}=\frac{1}{4\pi^{2}\tau^{4}}\ \ \ ;\ \ \ C_{5}=-\frac{1}{4\pi^{2}\tau^{4}}
\end{equation}
It is obvious that these coefficients satisfy the traceless condition in Eq.~(\ref{conformal}) with $N=2$. For $N>2$, with general $\xi$,
\begin{eqnarray}
C_{1}&=&\frac{\Gamma\left[\frac{N}{2}+1\right]\Gamma\left[\frac{N}{2}-1\right]}{2\pi^{N}\tau^{2N}}
\left[N(N-2)-8(N^2-N-2)\xi+16(N^2-1)\xi^{2}\right]\\
C_{2}&=&-\frac{\Gamma\left[\frac{N}{2}+1\right]\Gamma\left[\frac{N}{2}-1\right]}{4\pi^{N}\tau^{2N}}
\left[(N-2)-4(N-1)\xi\right]^{2}\\
C_{3}&=&\frac{\Gamma\left[\frac{N}{2}+1\right]\Gamma\left[\frac{N}{2}-1\right]}{4\pi^{N}\tau^{2N}}
\left[(N-2)-8(N-2)\xi+16(N-1)\xi^{2}\right]\\
C_{4}&=&\frac{\Gamma\left[\frac{N}{2}\right]\Gamma\left[\frac{N}{2}-1\right]}{8\pi^{N}\tau^{2N}}
\left[(N-2)-4(N-2)\xi+8(N-1)\xi^{2}\right]\\
C_{5}&=&\frac{\Gamma\left[\frac{N}{2}\right]\Gamma\left[\frac{N}{2}-1\right]}{16\pi^{N}\tau^{2N}}
\left[(N-2)(N^2-N-4)-8(N-2)(N^2-3)\xi\right.\nonumber\\
&&\ \ \ \ \ \ \ \ \ \ \ \left.+16(N-1)(N^2-3)\xi^2\right]\nonumber\\
\end{eqnarray}
Hence,
\begin{eqnarray}
C_{1}+NC_{2}-4C_{3}&=&\frac{4\Gamma\left[\frac{N}{2}+1\right]\Gamma\left[\frac{N}{2}-1\right]}{\pi^{N}\tau^{2N}}(N-2)(N-1)^{2}
\left[\xi-\frac{(N-2)}{4(N-1)}\right]^{2}\\
C_{2}+2C_{4}+NC_{5}&=&\frac{2\Gamma^{2}\left[\frac{N}{2}\right]}{\pi^{N}\tau^{2N}}(N+1)(N-1)^{2}
\left[\xi-\frac{(N-2)}{4(N-1)}\right]^{2}
\end{eqnarray}
They vanish when $\xi$ is equal to the conformal value $(N-2)/4(N-1)$.

The correlation of the stress-energy tensor has been considered in \cite{MarVer00} in relation to the theory of stochastic gravity \cite{HuVer08}. Here we have given the correlation explicitly in terms of the modified Bessel functions and we have also extended the result to arbitrary dimensions.


\subsection{Stress-energy Tensor Correlators in Euclidean $AdS^N$ space}

To obtain the five coefficients $C_{1}(\tau)$ to $C_{5}(\tau)$ on $H^{N}$, we just need to evaluate five components of $\Delta T_{\mu\nu\alpha'\beta'}^{2}(x,x')$. Here we shall choose $\Delta T_{\sigma\sigma\sigma'\sigma'}^{2}(x,x')$, $\Delta T_{\sigma\sigma\theta'\theta'}^{2}(x,x')$, $\Delta T_{\sigma\theta\sigma'\theta'}^{2}(x,x')$, $\Delta T_{\theta\theta\phi'_{1}\phi'_{1}}^{2}(x,x')$ and $\Delta T_{\theta\phi_{1}\theta'\phi'_{1}}^{2}(x,x')$. Since $H^{N}$ is homogeneous, the correlator should only depend on the geodesic distance between the two points. Hence, it is possible to simplify the consideration by appropriately choosing $x$ and $x'$. First, we shall set $x$ and $x'$ to have the same angular coordinates $\Omega'\rightarrow\Omega$. After that we shall take the limit $\sigma'\rightarrow 0$. In effect we shall take $x'$ to be at the origin.

As $\Omega'\rightarrow\Omega$, various bitensors can be simplified as follows.
\begin{eqnarray}
\tau(x,x')&=&|\sigma-\sigma'|,\\
n_{\mu}(x,x')&=&\delta_{\mu\sigma},\\
n_{\alpha'}(x,x')&=&-\delta_{\alpha'\sigma'},\\
g_{\sigma\sigma'}(x,x')&=&1,\\
g_{\theta\theta'}(x,x')&=&a^{2}\sinh\left(\frac{\sigma}{a}\right)\sinh\left(\frac{\sigma'}{a}\right),\\
g_{\phi_{1}\phi'_{1}}(x,x')&=&a^{2}\sinh\left(\frac{\sigma}{a}\right)\sinh\left(\frac{\sigma'}{a}\right)\sin^{2}\theta,
\end{eqnarray}
and so on. Note that the non-diagonal elements of $g_{\mu\alpha'}$ vanish in this limit. Using this result the relationship between the various components of $\Delta T_{\mu\nu\alpha'\beta'}^{2}(x,x')$ and the coefficients $C_{i}$ also simplify. Then we have, as $\Omega'\rightarrow\Omega$,
\begin{eqnarray}
\Delta T_{\sigma\sigma\sigma'\sigma'}^{2}(x,x')&=&C_{1}+2C_{2}-4C_{3}+2C_{4}+C_{5},\\
\Delta T_{\sigma\sigma\theta'\theta'}^{2}(x,x')&=&a^{2}\sinh^{2}\left(\frac{\sigma'}{a}\right)(C_{2}+C_{5}),\\
\Delta T_{\sigma\theta\sigma'\theta'}^{2}(x,x')&=&-a^{2}\sinh\left(\frac{\sigma}{a}\right)\sinh\left(\frac{\sigma'}{a}\right)(C_{3}-C_{4}),\\
\Delta T_{\theta\theta\phi'_{1}\phi'_{1}}^{2}(x,x')&=&a^{4}\sinh^{2}\left(\frac{\sigma}{a}\right)\sinh^{2}\left(\frac{\sigma'}{a}\right)\sin^{2}\theta C_{5},\\
\Delta T_{\theta\phi_{1}\theta'\phi'_{1}}^{2}(x,x')&=&a^{4}\sinh^{2}\left(\frac{\sigma}{a}\right)\sinh^{2}\left(\frac{\sigma'}{a}\right)\sin^{2}\theta C_{4},\label{T2comp5}
\end{eqnarray}
The evaluation of the various components of the correlator can be further simplified if we take $\sigma'\rightarrow 0$. As we have seen in the last section, only terms with low values of $l$ and $l'$ will contribute.

To see how the procedure goes, we consider $\Delta T_{\theta\phi_{1}\theta'\phi'_{1}}^{2}(x,x')$ in some details. Using the prescription in Eq.~(\ref{T2finalexp}),
\begin{eqnarray}
\Delta T_{\theta\phi_{1}\theta'\phi'_{1}}^{2}(x,x')&=&\frac{1}{2}
\int_{0}^{\infty}du u^{\nu}\int_{0}^{\infty}dv v^{\nu}\int_{0}^{\infty}d\kappa\int_{0}^{\infty}d\kappa'\sum_{lml'm'}
e^{-\frac{u}{a^{2}}(\kappa_{2}+a^{2}b)}e^{-\frac{v}{a^{2}}(\kappa'_{2}+a^{2}b)}\nonumber\\
&&\ \ \ \ \ \ \ \ T_{\theta\phi_{1}}[\phi_{\kappa lm}(x),\phi_{\kappa' l'm'}^{*}(x)]T_{\theta'\phi'_{1}}[\phi_{\kappa' l'm'}(x'),\phi_{\kappa lm}^{*}(x')]
\end{eqnarray}
To anticipate that we shall take $\sigma'\rightarrow 0$, we only need to consider the first few values of $l$ and $l'$. For example, when $l=l'=1$, we have
\begin{eqnarray}
T_{\theta\phi_{1}}[\phi_{\kappa 1m}(x),\phi_{\kappa' 1m'}(x)]&=&
-f_{\kappa 1}(\sigma)f_{\kappa'1}(\sigma)\left[\partial_{\theta}Y_{1m'}^{*}(\Omega)\partial_{\phi_{1}}Y_{1m}(\Omega)
+\partial_{\phi_{1}}Y_{1m'}^{*}(\Omega)\partial_{\theta}Y_{1m}(\Omega)\right.\nonumber\\
&&\ \ \ \ \ \ \ \ \ \ \ \ \ \ \ \ \ \ \ \ \ \ \left.-2\xi\nabla_{\theta}\nabla_{\phi_{1}}\left(Y_{1m'}^{*}(\Omega)Y_{1m}(\Omega)\right)\right]
\end{eqnarray}
The summations over $m$ and $m'$ can be carried out using the addition theorem in Eq.~(\ref{addtheorem}). Moreover, in the limit $\Omega'\rightarrow\Omega$, we have
\begin{eqnarray}
\left.\partial_{\theta}\partial_{\theta'}(\Omega\cdot\Omega')\right|_{\Omega'\rightarrow\Omega}&=&1\\
\left.\partial_{\phi_{1}}\partial_{\phi'_{1}}(\Omega\cdot\Omega')\right|_{\Omega'\rightarrow\Omega}&=&\sin^{2}\theta\\
\left.\partial_{\phi_{1}}\partial_{\phi'_{1}}(\Omega\cdot\Omega')^{2}\right|_{\Omega'\rightarrow\Omega}&=&2\sin^{2}\theta\\
\left.\partial_{\theta}\partial_{\phi_{1}}\partial_{\phi'_{1}}(\Omega\cdot\Omega')^{2}\right|_{\Omega'\rightarrow\Omega}&=&2\sin\theta\cos\theta\\
\left.\partial_{\theta}\partial_{\theta'}\partial_{\phi_{1}}\partial_{\phi'_{1}}(\Omega\cdot\Omega')^{2}\right|_{\Omega'\rightarrow\Omega}&=&2
\end{eqnarray}
and the summations can be simplified to
\begin{eqnarray}
&&\left.\sum_{mm'}T_{\theta\phi_{1}}[\phi_{\kappa 1m}(x),\phi_{\kappa' 1m'}^{*}(x)]T_{\theta'\phi'_{1}}[\phi_{\kappa' 1m'}(x'),\phi_{\kappa 1m}^{*}(x')]\right|_{\Omega'\rightarrow\Omega}\nonumber\\
&=&\frac{2\Gamma^{2}\!\left(\frac{N}{2}+1\right)(1-2\xi)^{2}}{\pi^{N}}\sin^{2}\theta \left[f_{\kappa 1}(\sigma)f_{\kappa'1}(\sigma)f_{\kappa 1}(\sigma')f_{\kappa'1}(\sigma')\right]
\end{eqnarray}
Furthermore, as $\sigma'\rightarrow 0$,
\begin{eqnarray}
&&\left.\sum_{mm'}T_{\theta\phi_{1}}[\phi_{\kappa 1m}(x),\phi_{\kappa' 1m'}^{*}(x)]T_{\theta'\phi'_{1}}[\phi_{\kappa' 1m'}(x'),\phi_{\kappa 1m}^{*}(x')]\right|_{\Omega'\rightarrow\Omega,\sigma'\rightarrow 0}\nonumber\\
&=&\frac{2(1-2\xi)^{2}}{(2\pi)^{N}}\sinh^{2}\left(\frac{\sigma'}{a}\right)\sin^{2}\theta\left[c_{1}(\kappa)f_{\kappa 1}(\sigma)c_{1}(\kappa')f_{\kappa'1}(\sigma)\right]
\end{eqnarray}
where we see that $\sinh^{2}(\sigma'/a)$ is the leading behavior as $\sigma'\rightarrow 0$. This behavior conforms with that of $\Delta T_{\theta\phi_{1}\theta'\phi'_{1}}^{2}(x,x')$ in Eq.~(\ref{T2comp5}). The terms with the same leading behavior are with $(l,l')=(0,2)$ and $(2,0)$. Following the same method as above, we have
\begin{eqnarray}
&&\left.\sum_{mm'}T_{\theta\phi_{1}}[\phi_{\kappa 0m}(x),\phi_{\kappa' 2m'}^{*}(x)]T_{\theta'\phi'_{1}}[\phi_{\kappa' 2m'}(x'),\phi_{\kappa 0m}^{*}(x')]\right|_{\Omega'\rightarrow\Omega,\sigma'\rightarrow 0}\nonumber\\
&=&\frac{4\xi^{2}}{(2\pi)^{N}}\sinh^{2}\left(\frac{\sigma'}{a}\right)\sin^{2}\theta\left[c_{0}(\kappa)f_{\kappa 0}(\sigma)c_{2}(\kappa')f_{\kappa'2}(\sigma)\right]\\
&&\left.\sum_{mm'}T_{\theta\phi_{1}}[\phi_{\kappa 2m}(x),\phi_{\kappa' 0m'}^{*}(x)]T_{\theta'\phi'_{1}}[\phi_{\kappa' 0m'}(x'),\phi_{\kappa 2m}^{*}(x')]\right|_{\Omega'\rightarrow\Omega,\sigma'\rightarrow 0}\nonumber\\
&=&\frac{4\xi^{2}}{(2\pi)^{N}}\sinh^{2}\left(\frac{\sigma'}{a}\right)\sin^{2}\theta\left[c_{2}(\kappa)f_{\kappa 2}(\sigma)c_{0}(\kappa')f_{\kappa'0}(\sigma)\right]
\end{eqnarray}
Comparing with Eq.(\ref{T2comp5}), we can extract the coefficient $C_{4}$.
\begin{eqnarray}
C_{4}(\sigma)=\frac{1}{(2\pi)^{N}a^{4}\left(\sinh\frac{\sigma}{a}\right)^{2}}\left[(1-2\xi)^{2}I_{1}^{(0)}I_{1}^{(0)}
+4\xi^{2}I_{0}^{(0)}I_{2}^{(0)}\right]
\end{eqnarray}
where we have defined the integral,
\begin{eqnarray}
I_{l}^{(i)}=\int_{0}^{\infty}du u^{\nu}\int_{0}^{\infty}d\kappa\kappa^{2i} e^{-\frac{u}{a^{2}}(\kappa^{2}+a^{2}b)}c_{l}(\kappa)f_{\kappa l}(\sigma)\label{Iintegral}
\end{eqnarray}

After some straight forward but somewhat lengthy calculations similar to that above, we have
\begin{eqnarray}
C_{1}&=&\frac{(N-1)\left[N(N-1)-4\xi(N^{2}+1)\right]}{4(2\pi)^{N}a^{4}N}\times\nonumber\\
&&\ \ \ \ \ \left[\left(a^{2}\xi\partial_{\sigma}\partial_{\sigma}-a\xi
\left(\coth\frac{\sigma}{a}\right)\partial_{\sigma}\right)\left(I_{0}^{(0)}\right)^{2}
-a^{2}\left(\partial_{\sigma}I_{0}^{(0)}\right)^{2}\right]\nonumber\\
&&+\frac{N-4\xi(N-1)}{(2\pi)^{N}a^{4}N}
\left[\left(a^{2}\xi\partial_{\sigma}\partial_{\sigma}-a\xi\left(\coth\frac{\sigma}{a}\right)
\partial_{\sigma}\right)\left(I_{0}^{(0)}I_{0}^{(1)}\right)
-a^{2}\left(\partial_{\sigma}I_{0}^{(0)}\right)\left(\partial_{\sigma}I_{0}^{(1)}\right)\right]\nonumber\\
&&+\frac{1}{(2\pi)^{N}a^{4}}\left[2a^{2}(1-\xi-4\xi^{2})\left(\partial_{\sigma}I_{1}^{(0)}\right)^{2}
+(1-2\xi)(1-4\xi)\left(I_{1}^{(0)}I_{1}^{(1)}\right)\right]\nonumber\\
&&-\frac{1}{(2\pi)^{N}a^{4}}\left[a^{2}\xi(3-8\xi)\partial_{\sigma}\partial_{\sigma}
+\frac{a(1-4\xi)}{\sinh\frac{\sigma}{a}}\left[2(1-2\xi)-\xi\cosh\frac{\sigma}{a}\right]
\partial_{\sigma}\right.\nonumber\\
&&\ \ \ \ \ \ \ \ \ \ -\frac{(N-1)(1-2\xi)}{4}\left((N-1)-4\xi(N+1)\right)\nonumber\\
&&\ \ \ \ \ \ \ \ \ \ \left.-\frac{2}{\left(\sinh\frac{\sigma}{a}\right)^{2}}
\left((1-5\xi+8\xi^{2})-4\xi(1-2\xi)\left(\cosh\frac{\sigma}{a}\right)\right)\right]
\left(I_{1}^{(0)}\right)^{2}\nonumber\\
&&-\frac{1}{(2\pi)^{N}a^{4}}\left[\frac{2a^{2}\xi}{N}((N+2)+4N\xi)\left(\partial_{\sigma}I_{0}^{(0)}\right)
\left(\partial_{\sigma}I_{2}^{(0)}\right)\right.\nonumber\\
&&\ \ \ \ \ -\left.\frac{8a\xi}{\sinh\frac{\sigma}{a}}\left(\partial_{\sigma}I_{0}^{(0)}\right)I_{2}^{(0)}
+\xi(1-4\xi)\left(I_{0}^{(1)}I_{2}^{(0)}+I_{0}^{(0)}I_{2}^{(1)}\right)\right]\nonumber
\end{eqnarray}
\begin{eqnarray}
&&+\frac{1}{(2\pi)^{N}a^{4}}\left[\frac{4a^{2}\xi^{2}(N+1)}{N}\partial_{\sigma}\partial_{\sigma}
-\frac{4a\xi^{2}}{N\sinh\frac{\sigma}{a}}\left[4N+\cosh\frac{\sigma}{a}\right]\partial_{\sigma}\right.\nonumber\\
&&\ \ \ \ \ \ \ \ \ \ \left.-\frac{N-1}{2}\left((N-1)\xi-4\xi^{2}(N+1)\right)+\frac{8\xi^{2}}{\left(\sinh\frac{\sigma}{a}\right)^{2}}
\left(1+2\cosh\frac{\sigma}{a}\right)\right]\left(I_{0}^{(0)}I_{2}^{(0)}\right)\nonumber\\  \\
C_{2}&=&-\frac{(N-1)\left[N(N-1)-4\xi(N^{2}+1)\right]}{4(2\pi)^{N}a^{4}N}\times\nonumber\\
&&\ \ \ \ \ \left[\left(a^{2}\xi\partial_{\sigma}\partial_{\sigma}
-a\xi\left(\coth\frac{\sigma}{a}\right)\partial_{\sigma}\right)\left(I_{0}^{(0)}\right)^{2}
-a^{2}\left(\partial_{\sigma}I_{0}^{(0)}\right)^{2}\right]\nonumber\\
&&-\frac{N-4\xi(N-1)}{(2\pi)^{N}a^{4}N}\left[\left(a^{2}\xi\partial_{\sigma}\partial_{\sigma}
-a\xi\left(\coth\frac{\sigma}{a}\right)
\partial_{\sigma}\right)\left(I_{0}^{(0)}I_{0}^{(1)}\right)
-a^{2}\left(\partial_{\sigma}I_{0}^{(0)}\right)\left(\partial_{\sigma}I_{0}^{(1)}\right)\right]\nonumber\\
&&+\frac{1-4\xi}{(2\pi)^{N}a^{4}}\left[\left(a^{2}\xi\partial_{\sigma}\partial_{\sigma}
-a\xi\left(\coth\frac{\sigma}{a}\right)\partial_{\sigma}
+\frac{1}{\left(\sinh\frac{\sigma}{a}\right)^{2}}\right)\left(I_{1}^{(0)}\right)^{2}
-a^{2}\left(\partial_{\sigma}I_{1}^{(0)}\right)^{2}\right]\nonumber\\
&&-\frac{4\xi}{(2\pi)^{N}a^{4}N}\left[\left(a^{2}\xi\partial_{\sigma}\partial_{\sigma}
-a\xi\left(\coth\frac{\sigma}{a}\right)\partial_{\sigma}\right)
\left(I_{0}^{(0)}I_{2}^{(0)}\right)
-a^{2}\left(\partial_{\sigma}I_{0}^{(0)}\right)\left(\partial_{\sigma}I_{2}^{(0)}\right)\right]\label{Coeff1}\\
C_{3}&=&-\frac{1-2\xi}{2(2\pi)^{N}a^{4}\left(\sinh\frac{\sigma}{a}\right)^{2}}
\left[a(1-4\xi)\left(\sinh\frac{\sigma}{a}\right)\partial_{\sigma}-2(1-2\xi)
+4\xi\left(\cosh\frac{\sigma}{a}\right)\right]
\left(I_{1}^{(0)}\right)^{2}\nonumber\\
&&-\frac{2\xi}{(2\pi)^{N}a^{4}\left(\sinh\frac{\sigma}{a}\right)^{2}}
\left[2\xi\left(a\left(\sinh\frac{\sigma}{a}\right)\partial_{\sigma}
-\left(1+\cosh\frac{\sigma}{a}\right)\right)\left(I_{0}^{(0)}I_{2}^{(0)}\right)\right.\nonumber\\
&&\ \ \ \ \ \ \ \ \ \ \ \ \ \ \ \ \ \left.-a\left(\sinh\frac{\sigma}{a}\right)\left(\partial_{\sigma}I_{0}^{(0)}\right)\left(I_{2}^{(0)}\right)\right]\\
C_{4}&=&\frac{(1-2\xi)^{2}}{(2\pi)^{N}a^{4}\left(\sinh\frac{\sigma}{a}\right)^{2}}
\left(I_{1}^{(0)}\right)^{2}+\frac{4\xi^{2}}{(2\pi)^{N}a^{4}
\left(\sinh\frac{\sigma}{a}\right)^{2}}\left(I_{0}^{(0)}I_{2}^{(0)}\right)\\
C_{5}&=&-\frac{(1-4\xi)}{2(2\pi)^{N}a^{4}N}\bigg[a^{2}\left(N-4\xi(N-1)\right)\left(\partial_{\sigma}I_{0}^{(0)}\right)
\left(\partial_{\sigma}I_{0}^{(2)}\right)\nonumber\\
&&\ \ \ \ \ \ \ \ \ \ \left.+\frac{a^{2}(N-1)}{4}\left(N(N-1)-4\xi(N^{2}+1)\right)
\left(\partial_{\sigma}I_{0}^{(0)}\right)^{2}\right]\nonumber\\
&&+\frac{1}{2(2\pi)^{N}a^{4}N}\left[\left(N-2\xi(2N-1)+8\xi^{2}(N-1)\right)\left(I_{0}^{(1)}\right)^{2}\right.\nonumber\\
&&\ \ \ \ \ \ \ \ \ \ \left.-2\left(\xi(2N-1)-4\xi^{2}(N-1)\right)\left(I_{0}^{(0)}I_{0}^{(2)}\right)\right]\nonumber\\
&&-\frac{1}{4(2\pi)^{N}a^{4}N}\left[4a\xi \left(\coth\frac{\sigma}{a}\right)\left(N-4\xi(N-1)\right)\partial_{\sigma}\right.\nonumber\\
&&\ \ \ \ \ \ \ \ \ \ \ \ \ \left.-(N-1)\left(N(N-1)-4\xi(2N^{2}-N+1)+16\xi^{2}N^{2}\right)\right]
\left(I_{0}^{(0)}I_{0}^{(1)}\right)\nonumber\\
&&+\frac{(N-1)}{32(2\pi)^{N}a^{4}N}\left[-8a\xi \left(\coth\frac{\sigma}{a}\right)\left(N(N-1)-4\xi(N^{2}+1)\right)\partial_{\sigma}\right.\nonumber\\
&&+(N-1)\left(N(N-1)^{2}-4\xi(N-1)(2N^{2}+N+1)+16\xi^{2}(N+1)(N^{2}+1)\right)\left]
\left(I_{0}^{(0)}\right)^{2}\right.\nonumber\\
&&+\frac{(1-4\xi)^{2}}{2(2\pi)^{N}a^{4}}\left[a^{2}
\left(\partial_{\sigma}I_{1}^{(0)}\right)\left(\partial_{\sigma}I_{1}^{(0)}\right)
-I_{1}^{(0)}I_{1}^{(1)}\right]\nonumber
\end{eqnarray}
\begin{eqnarray}
&&+\frac{(1-4\xi)}{2(2\pi)^{N}a^{4}}\left[2a\xi\left(\coth\frac{\sigma}{a}\right)\partial_{\sigma}
-\frac{N-1}{4}\left((N-1)-4\xi(N+1)\right)\right.\nonumber\\
&&\ \left.+\frac{1}{\left(\sinh\frac{\sigma}{a}\right)^{2}}
\left((N-5)-4\xi(N-2)\right)\right]\left(I_{1}^{(0)}\right)^{2}\nonumber\\
&&+\frac{\xi(1-4\xi)}{(2\pi)^{N}a^{4}N}\left[\left(I_{0}^{(0)}I_{2}^{(1)}+I_{0}^{(1)}I_{2}^{(0)}\right)
-2a^{2}\left(\partial_{\sigma}I_{0}^{(0)}\right)\left(\partial_{\sigma}I_{2}^{(0)}\right)\right]\nonumber\\
&&-\frac{1}{2(2\pi)^{N}a^{4}N}\left[8a\xi^{2}\left(\coth\frac{\sigma}{a}\right)\partial_{\sigma}-(N-1)
\left((N-1)\xi-4\xi^{2}(N+1)\right)\right]\left(I_{0}^{(0)}I_{2}^{(0)}\right)\nonumber\\ \label{Coeff5}
\end{eqnarray}

Although we have succeeded in expressing the coefficients in terms of products of integrals involving the associated Legendre functions, the expressions are a bit complicated and thus are not very illuminating. Moreover, since the integrals cannot be simplified to known functions, it is hard to check the corresponding conservation conditions and the traceless condition in the conformally coupled cases. Therefore, we shall consider in the next section the asymptotic behaviors of the above coefficients in the small and large geodesic distance limits. In so doing we could have a better understanding of the correlators and we could also check the conservation and the traceless conditions explicitly for various dimensions.

%
%

\section{Small and large geodesic distance limits in AdS for different dimensions}\label{sec:limits}


In this section we explore the small and large geodesic distance limits of the coefficients $C_{1}$ to $C_{5}$ in Euclidean $AdS^N$ for arbitrary $N$.  As we have mentioned above, it is possible to check the conservation and the traceless conditions explicitly for various dimensions in the small distance limit. In addition, the small distance limit also indicates the divergent behavior of the correlators in the coincident limit as $\sigma\rightarrow 0$. This type of expansion has been explored in \cite{OsbSho00} in relation to the generalization of the $c$-theorem \cite{Zam86} from two to higher dimensions. A general discussion has also been given in \cite{OsbSho00} on the stress-energy correlators of conformally coupled scalar and fermion fields in constant curvature spaces.

For the small distance limit, we need to consider the integral $I_{l}^{(i)}(N)$ in Eq.~(\ref{Iintegral}) in more detail. First we shall use the integral representation of the Legendre function,
\begin{equation}
P_{-\frac{1}{2}+i\kappa}^{1-l-\frac{N}{2}}\left(\cosh\frac{\sigma}{a}\right)=
\frac{\sqrt{2}\left(\sinh\frac{\sigma}{a}\right)^{1-l-\frac{N}{2}}}{\sqrt{\pi}\Gamma\left(-\frac{1}{2}+l+\frac{N}{2}\right)}
\left(\frac{\sigma}{a}\right)\int_{0}^{1}dt\ \!\cos\left(\frac{\kappa\sigma t}{a}\right)\left(\cosh\frac{\sigma}{a}-\cosh\frac{\sigma t}{a}\right)^{l+\frac{N}{2}-\frac{3}{2}}
\end{equation}
Since the normalization constant is given by
\begin{equation}
|c_{l}(\kappa)|^{2}=\left[\kappa^{2}+\left(\rho_{N}+l-1\right)^{2}\right]\left[\kappa^{2}+\left(\rho_{N}+l-2\right)^{2}\right]
\cdots\left[\kappa^{2}+\left(\rho_{N}\right)^{2}\right]|c_{0}(\kappa)|^{2}
\end{equation}
and for even dimensions, $|c_{0}(\kappa)|^{2}$ is given by Eq.~(\ref{c0even}),
we see that if we want to extract the leading contribution of the integrals as $\sigma\rightarrow 0$, we need to deal with integrals of the general form
\begin{eqnarray}
&&\int_{0}^{1}dt\ \!\left(\cosh\frac{\sigma}{a}-\cosh\frac{\sigma t}{a}\right)^{l+\frac{N}{2}-\frac{3}{2}}\int_{0}^{\infty}du\ \! u^{\nu}\int_{0}^{\infty}d\kappa\ \! \kappa^{2n+2i+1}\left(\cos\frac{\kappa\sigma t}{a}\right)\ \! e^{-\frac{u}{a^{2}}(\kappa^{2}+a^{2}b)}\nonumber\\
&=&\int_{0}^{1}dt\ \!\left(\cosh\frac{\sigma}{a}-\cosh\frac{\sigma t}{a}\right)^{l+\frac{N}{2}-\frac{3}{2}}
\int_{0}^{\infty}d\kappa\frac{a^{2(\nu+1)}\Gamma(\nu+1)\kappa^{2n+2i+1}\left(\cos\frac{\kappa\sigma t}{a}\right)}{(\kappa^{2}+a^{2}b)^{\nu+1}}
\end{eqnarray}
where $n$ is an integer. This is finite for sufficiently large values of $\nu$. In this analytic continuation procedure, we can represent $\kappa^{2n+2i}$ by derivatives on the cosine function. Then the above integral becomes
\begin{equation}
\int_{0}^{1}dt\ \!\left(\cosh\frac{\sigma}{a}-\cosh\frac{\sigma t}{a}\right)^{l+\frac{N}{2}-\frac{3}{2}}
(-1)^{n+i}\left(\frac{\partial}{\partial\eta}\right)^{2n+2i}\int_{0}^{\infty}d\kappa\left(\frac{a^{2}\kappa\ \!\cos\eta\kappa}{\kappa^{2}+a^{2}b}\right)
\end{equation}
where $\eta=\sigma t/a$. Since the integration over $\kappa$ is now finite even when $\nu\rightarrow 0$, we have taken that limit.

To see how it goes explicitly, we consider the case with $l=1$, $N=4$ and $n=i=1$. Then, expanding in powers of $\sigma/a$,
\begin{eqnarray}
&&\left(\cosh\frac{\sigma}{a}-\cosh\frac{\sigma t}{a}\right)^{3/2}\left(\frac{\partial}{\partial\eta}\right)^{4}\int_{0}^{\infty}d\kappa\left(\frac{a^{2}\kappa\ \!\cos\eta\kappa}{\kappa^{2}+a^{2}b}\right)\nonumber\\
&=&a^{2}\left(\frac{a}{\sigma}\right)\left(\frac{3}{\sqrt{2}}\right)(1-t^2)^{3/2}t^{-4}+
a^{2}\left(\frac{\sigma}{a}\right)\left(\frac{1}{8\sqrt{2}}\right)(1-t^{2})^{3/2}\left[3t^{-4}+(3+4a^{2}b)t^{-2}\right]\nonumber\\
&&\ \ \ +a^{2}\left(\frac{\sigma}{a}\right)^{3}\left(\frac{1}{640\sqrt{2}}\right)(1-t^{2})^{3/2}\Big\{13t^{-4}+2(9+20a^{2}b)t^{-2}
+13+40a^{2}b\nonumber\\
&&\ \ \ \ \ \ \ \ \ \ \ \ \left.-160a^{4}b^{2}\left[2\gamma+\ \!{\rm ln}\left[\left(\frac{\sigma}{a}\right)^{2}\left(a^{2}bt\right)\right]\right]\right\}
\end{eqnarray}
Using the formula,
\begin{equation}
\int_{0}^{1}dt\ \!t^{\alpha}(1-t^2)^{\beta}=\frac{\Gamma\left(\frac{1+\alpha}{2}\right)
\Gamma\left(1+\beta\right)}{2\Gamma\left(\frac{3}{2}+\frac{\alpha}{2}+\beta\right)}
\end{equation}
the integration over $t$ can be performed giving
\begin{eqnarray}
&&\int_{0}^{1}dt\ \!\left(\cosh\frac{\sigma}{a}-\cosh\frac{\sigma t}{a}\right)^{3/2}\left(\frac{\partial}{\partial\eta}\right)^{4}\int_{0}^{\infty}d\kappa\left(\frac{\kappa\ \!\cos\eta\kappa}{\kappa^{2}+a^{2}b}\right)\nonumber\\
&=&a^{2}\left(\frac{a}{\sigma}\right)\left(\frac{3\pi}{2\sqrt{2}}\right)
-a^{2}\left(\frac{\sigma}{a}\right)\left(\frac{3\pi}{32\sqrt{2}}\right)
(1+4a^{2}b)-a^{2}\left(\frac{\sigma}{a}\right)^{3}\left(\frac{\pi}{10240\sqrt{2}}\right)\times\nonumber\\
&&\ \ \left\{73+360a^{2}b-240(a^{2}b)^{2}
\left[3-4\gamma-2\ \!{\rm ln}\left(\frac{a^{2}b}{4}\right)-4\ \!{\rm ln}\left(\frac{\sigma}{a}\right)\right]\right\}
\end{eqnarray}

Since the coefficients $C_{1}$ to $C_{5}$ are expressed in terms of the integrals $I_{l}^{(i)}(N)$, the expansion above can be used to obtain the corresponding small geodesic distance limits of the coefficients.
For $N=4$,
\begin{eqnarray}
C_{1}(4)&=&\frac{8}{\pi^{4}\sigma^{8}}(1-10\xi+30\xi^{2})\nonumber\\
&&-\frac{1}{6\pi^{4}a^{2}\sigma^{6}}\bigg\{(1+20\xi+81\xi^{2})+72\xi^{2}\left(\gamma+\psi\left(\frac{1}{2}+a\sqrt{b}\right)+{\rm ln}\left(\frac{\sigma}{2a}\right)\right)\nonumber\\
&&\ \ \ \ \ \ \ +12a^{2}b\left[(1-12\xi+37\xi^{2})-24\xi^{2}\left(\gamma+\psi\left(\frac{1}{2}+a\sqrt{b}\right)+{\rm ln}\left(\frac{\sigma}{2a}\right)\right)\right]\bigg\}+\cdots\nonumber\\
C_{2}(4)&=&-\frac{2}{\pi^{4}\sigma^{8}}(1-6\xi)^{2}\nonumber\\
&&+\frac{1}{24\pi^{4}a^{2}\sigma^{6}}\bigg\{(10-102\xi+345\xi^{2})+36\xi^{2}\left(\gamma+\psi\left(\frac{1}{2}+a\sqrt{b}\right)+{\rm ln}\left(\frac{\sigma}{2a}\right)\right)\nonumber\\
&&\ \ \ \ \ \ \ +12a^{2}b\left[(-6+29\xi^{2})-12\xi^{2}\left(\gamma+\psi\left(\frac{1}{2}+a\sqrt{b}\right)+{\rm ln}\left(\frac{\sigma}{2a}\right)\right)\right]\bigg\}+\cdots\nonumber\\
C_{3}(4)&=&\frac{1}{\pi^{4}\sigma^{8}}(1-8\xi+24\xi^{2})\nonumber\\
&&-\frac{1}{96\pi^{4}a^{2}\sigma^{6}}\bigg\{(11-40\xi+420\xi^{2})+144\xi^{2}\left(\gamma+\psi\left(\frac{1}{2}+a\sqrt{b}\right)+{\rm ln}\left(\frac{\sigma}{2a}\right)\right)\nonumber\\
&&\ \ \ \ \ \ \ +12a^{2}b\left[(3-24\xi+68\xi^{2})-576\xi^{2}\left(\gamma+\psi\left(\frac{1}{2}+a\sqrt{b}\right)+{\rm ln}\left(\frac{\sigma}{2a}\right)\right)\right]\bigg\}+\cdots\nonumber\\
C_{4}(4)&=&\frac{1}{4\pi^{4}\sigma^{8}}(1-4\xi+12\xi^{2})\nonumber\\
&&+\frac{1}{96\pi^{4}a^{2}\sigma^{6}}\bigg\{(-5+20\xi-114\xi^{2})-24\xi^{2}\left(\gamma+\psi\left(\frac{1}{2}+a\sqrt{b}\right)+{\rm ln}\left(\frac{\sigma}{2a}\right)\right)\nonumber\\
&&\ \ \ \ \ \ \ +12a^{2}b\left[(-1+4\xi-10\xi^{2})+8\xi^{2}\left(\gamma+\psi\left(\frac{1}{2}+a\sqrt{b}\right)+{\rm ln}\left(\frac{\sigma}{2a}\right)\right)\right]\bigg\}+\cdots\nonumber\\
C_{5}(4)&=&\frac{1}{\pi^{4}\sigma^{8}}(1-13\xi+39\xi^{2})\nonumber\\
&&+\frac{1}{96\pi^{4}a^{2}\sigma^{6}}\bigg\{(-47+548\xi-1602\xi^{2})-24\xi^{2}\left(\gamma+\psi\left(\frac{1}{2}+a\sqrt{b}\right)+{\rm ln}\left(\frac{\sigma}{2a}\right)\right)\nonumber\\
&&\ \ \ \ \ \ \ +12a^{2}b\left[(1+4\xi-42\xi^{2})+8\xi^{2}\left(\gamma+\psi\left(\frac{1}{2}+a\sqrt{b}\right)+{\rm ln}\left(\frac{\sigma}{2a}\right)\right)\right]\bigg\}+\cdots\nonumber\\ \label{N4coeff}
\end{eqnarray}

For odd dimensions, $|c_{0}|^{2}$ is given by Eq.~(\ref{c0odd}) instead so one has to consider integrals of the general form,
\begin{eqnarray}
&&\int_{0}^{1}dt\ \!\left(\cosh\frac{\sigma}{a}-\cosh\frac{\sigma t}{a}\right)^{l+\frac{N}{2}-\frac{3}{2}}\int_{0}^{\infty}du\ \! u^{\nu}\int_{0}^{\infty}d\kappa\ \! \kappa^{2n+2i}\left(\cos\frac{\kappa\sigma t}{a}\right)\ \! e^{-\frac{u}{a^{2}}(\kappa^{2}+a^{2}b)}\nonumber
\end{eqnarray}
Similar procedure can be carried out as in the even dimensional cases to obtain an expansion in powers of $\sigma/a$. To be explicitly, we take $l=1$, $N=5$, $n=3$ and $i=1$ as an example:
\begin{eqnarray}
&&\int_{0}^{1}dt\ \!\left(\cosh\frac{\sigma}{a}-\cosh\frac{\sigma t}{a}\right)^{2}\int_{0}^{\infty}du\ \! u^{\nu}e^{-ub}\int_{0}^{\infty}d\kappa\ \! \kappa^{8}\left(\cos\frac{\kappa\sigma t}{a}\right)\ \! e^{-\frac{u}{a^{2}}\kappa^{2}}\nonumber\\
&=&\sqrt{\frac{\pi}{2}}a^{3/2}b^{-1/4}\int_{0}^{1}dt\ \!\left(\cosh\frac{\sigma}{a}-\cosh\frac{\sigma}{a}t\right)^{2}
\left(\frac{\partial}{\partial \eta}\right)^{8}\left[\eta^{\nu+\frac{1}{2}}K_{-\nu-\frac{1}{2}}\left(a\sqrt{b}\eta\right)\right]\nonumber\\
&=&\frac{\pi a^{2}}{2}\bigg\{-6(5+a^2 b)\frac{a}{\sigma}+\left[1+a^{2}b+(a^{2}b)^{2}\right]\frac{\sigma}{a}\nonumber\\
&&\ \ \ \ \ \ \ \ \ +\frac{1}{12}\left[1+a^{2}b+(a^{2}b)^{2}-3(a^{2}b)^{3}\right]\left(\frac{\sigma}{a}\right)^{3}+\cdots\bigg\}
\end{eqnarray}
where we have again taken $\eta=\sigma t/a$. Notice that we take the $\nu\rightarrow 0$ limit after the integration over $t$. Using this expansion one could obtain the small distance expansion for the coefficients $C_{1}$ to $C_{5}$ for odd dimensions.

Explicitly, for $N=5$,
\begin{eqnarray}
C_{1}(5)&=&\frac{45}{32 \pi^4 \sigma^{10}} (5 - 48 \xi + 128 \xi^2) -\frac{
 15}{32 \pi^4 a^2 \sigma^8} \left[(1 - 4 \xi )+ a^2 b (3 - 32 \xi + 128 \xi^2)\right]+\cdots\nonumber\\
C_{2}(5)&=&-\frac{15}{64 \pi^4 \sigma^{10}} (3 - 16 \xi)^2 + \frac{1}
{64 \pi^4 a^2  \sigma^8}\left[( 33 - 336 \xi + 832 \xi^2)+12\ \! a^2 b (1 - 8 \xi)^2 \right]+\cdots\nonumber\\
C_{3}(5)&=&\frac{15}{64 \pi^4 \sigma^{10}} (3 - 24 \xi + 64 \xi^2)\nonumber\\
&&\ \ \ \ \ -\frac{1}{64 \pi^4 a^2  \sigma^8}
\left[(9 - 54 \xi + 112 \xi^2) + 12 a^2 b (1 - 8 \xi + 32 \xi^2)\right]+\cdots\nonumber\\
C_{4}(5)&=&\frac{3}{64 \pi^4 \sigma^{10}}(3 - 12 \xi + 32 \xi^2)\nonumber\\
&&\ \ \ \ \ -\frac{1}{64 \pi^4 a^2  \sigma^8}
\left[(3 - 12 \xi + 28 \xi^2 )+3 a^2 b (1 - 4 \xi + 16 \xi^2)\right]+\cdots\nonumber\\
C_{5}(5)&=&\frac{3}{8 \pi^4 \sigma^{10}} (3 - 33 \xi + 88 \xi^2)\nonumber\\
&&\ \ \ \ \ -\frac{1}{64 \pi^4 a^2  \sigma^8}
\left[ 4 (9 - 96 \xi + 250 \xi^2)+3 a^2 b (1 - 24 \xi + 96 \xi^2)\right]+\cdots\label{N5coeff}
\end{eqnarray}

One can check that the conservation conditions in Eq.~(\ref{conservation}) are satisfied to the order indicated. For the traceless condition in Eq.~(\ref{conformal}), this is also true for the conformally coupled case. Here some clarification is needed. To the order indicated here it seems that the traceless conditions are satisfied when $\xi$ equals to the conformal value $(N-2)/4(N-1)$ while $b$ can take any value. However, this is not the case if we go higher orders. For example, going to one more order than Eq.~(\ref{N4coeff}) for $N=4$ and taking $\xi=1/6$, we have
\begin{eqnarray}
&&C_{1}+4C_{2}-4C_{3}\nonumber\\
&=&\frac{(1-4a^{2}b)}{384\pi^{4}a^{4}\sigma^{4}}\left\{(9-4a^{2}b)
+4(1-4a^{2}b)\left[\gamma+\psi\left(\frac{1}{2}+a\sqrt{b}\right)+{\rm ln}\left(\frac{\sigma}{2a}\right)\right]\right\}+\cdots\nonumber\\
&&C_{2}+2C_{4}+4C_{5}\nonumber\\
&=&-\frac{(1-4a^{2}b)}{768\pi^{4}a^{4}\sigma^{4}}\left\{(3+4a^{2}b)
+2(1-4a^{2}b)\left[\gamma+\psi\left(\frac{1}{2}+a\sqrt{b}\right)+{\rm ln}\left(\frac{\sigma}{2a}\right)\right]\right\}+\cdots\nonumber
\end{eqnarray}
Here $a^{2}b=m^{2}a^{2}+1/4$ for $N=4$. Therefore, the expressions above indeed vanish for $a^{2}b=1/4$ or $m=0$. The small geodesic distance limit for other dimensions, even or odd, can be carried out analogously as we have done above. We have listed the results in Appendix~\ref{appendix}.

With the expressions in the small distance limit of the coefficients $C$'s, it is possible to make some comparison with our results to that in \cite{PRVdS} and \cite{OsbSho00}. For simplicity, we just consider the leading behavior of these coefficients. In \cite{PRVdS}, these coefficients, named $P$, $Q$, $R$, $S$, and $T$ there, are expressed in terms of the Wightman function (in de Sitter spacetime),
\begin{eqnarray}
G(\sigma)=c_{m,N}F\left(h_{+},h_{-};\frac{N}{2};\frac{1+\cos\sigma/a}{2}\right)
\end{eqnarray}
where $F(\alpha,\beta;\gamma;z)$ is the hypergeometric function and
\begin{eqnarray}
h_{\pm}=\frac{1}{2}\left[(N-1)\pm\sqrt{(N-1)^{2}-4m^{2}a^{2}}\right]\ \ ;\ \ c_{m,N}=\frac{\Gamma(h_{+})\Gamma(h_{-})}{(4\pi a)^{N/2}\Gamma(N/2)}
\end{eqnarray}
In the small geodesic distance limit, $\sigma\rightarrow 0$,
\begin{eqnarray}
G(\sigma)=\frac{\Gamma\left(\frac{N}{2}-1\right)}{(4\pi)^{N/2}}\left(\frac{2}{\sigma}\right)^{N-2}+\cdots.
\end{eqnarray}
The leading behavior of the coefficients are then
\begin{eqnarray}
C_{1}(N)&=&\frac{N^{2}\Gamma^{2}(N/2)}{2\pi^{N}\sigma^{2N}}+\cdots\nonumber\\
C_{2}(N)&=&-\frac{N(N-2)\Gamma^{2}(N/2)}{4\pi^{N}\sigma^{2N}}+\cdots\nonumber\\
C_{3}(N)&=&\frac{N\Gamma^{2}(N/2)}{4\pi^{N}\sigma^{2N}}+\cdots\nonumber\\
C_{4}(N)&=&\frac{\Gamma^{2}(N/2)}{4\pi^{N}\sigma^{2N}}+\cdots\nonumber\\
C_{5}(N)&=&\frac{(N^{2}-N-4)\Gamma^{2}(N/2)}{8\pi^{N}\sigma^{2N}}+\cdots\label{PRVdScoeff}
\end{eqnarray}
After putting $\xi\rightarrow 0$, our results in Eqs.~(\ref{N4coeff}) and (\ref{N5coeff}) and also in Appendix~\ref{appendix} agree with Eq.~(\ref{PRVdScoeff}) above for $N>2$.

In \cite{OsbSho00} the conformally coupled case is considered. The coefficients, named $R_{-}$, $U_{-1}=U_{-2}$, $S_{-}$, $T_{-}$, and $V_{-}$ there, have the following leading behavior for small geodesic distance,
\begin{eqnarray}
C_{1}(N)&=&\frac{N\Gamma^{2}(N/2)}{(N-1)\pi^{N}\sigma^{2N}}+\cdots\nonumber\\
C_{2}(N)&=&0+\cdots\nonumber\\
C_{3}(N)&=&\frac{N\Gamma^{2}(N/2)}{4(N-1)\pi^{N}\sigma^{2N}}+\cdots\nonumber\\
C_{4}(N)&=&\frac{N\Gamma^{2}(N/2)}{8(N-1)\pi^{N}\sigma^{2N}}+\cdots\nonumber\\
C_{5}(N)&=&-\frac{\Gamma^{2}(N/2)}{4(N-1)\pi^{N}\sigma^{2N}}+\cdots\label{OScoeff}
\end{eqnarray}
From our results in Eqs.~(\ref{N4coeff}) and (\ref{N5coeff}) and also in Appendix~\ref{appendix}, since the leading terms do not depend on the mass $m$, we put $\xi$ to the conformal value $(N-2)/4(N-1)$ and obtain agreement with Eq.~(\ref{OScoeff}) above.

To consider the large geodesic distance limit for the various coefficients, we go back to the integral $I_{l}^{(i)}$ in Eq.~(\ref{Iintegral}). First,
\begin{equation}
f_{\kappa l}(\sigma)=c_{l}(\kappa)\left(\sinh\frac{\sigma}{a}\right)^{1-\frac{N}{2}}P_{-\frac{1}{2}+i\kappa}^{1-l-\frac{N}{2}}\left(\cosh\frac{\sigma}{a}\right)
\end{equation}
and from the asymptotic behavior as $\sigma\rightarrow\infty$ of the associated Legendre function, we have
\begin{equation}
P_{-\frac{1}{2}+i\kappa}^{1-l-\frac{N}{2}}\left(\cosh\frac{\sigma}{a}\right)\sim e^{-\frac{\sigma}{2a}}e^{i\frac{\kappa\sigma}{a}}
\end{equation}
The integrations over $u$ and $\kappa$ will only give powers of $\sigma$. Hence, the leading asymptotic behavior of $I_{l}^{(i)}$, independently of the parameters $l$ and $i$, is
\begin{equation}
I_{l}^{(i)}\sim e^{-(N-1)\sigma/2a}
\end{equation}
Using this result we can estimate the asymptotic behavior of the various coefficients in Eqs.~(\ref{Coeff1}) to (\ref{Coeff5}),
\begin{eqnarray}
C_{1},C_{2},C_{5}&\sim& e^{-(N-1)\sigma/a}\nonumber\\
C_{3}&\sim&e^{-N\sigma/a}\nonumber\\
C_{4}&\sim&e^{-(N+1)\sigma/a}\label{large}
\end{eqnarray}
This behavior is independent of the mass of the field. We can see that an intrinsic infrared cutoff scale is provided by the radius $a$ of the background spacetime.

This large geodesic distance limit can also be compared with the result in \cite{PRVdS}. Since there the authors considered the de Sitter spacetime, one could analytic continue their result by taking
\begin{equation}
Z=\cos\left(\frac{\sigma}{a}\right)\rightarrow\cosh\left(\frac{\sigma}{a}\right)\sim e^{\sigma/a}
\end{equation}
Then our result in Eq.~(\ref{large}) indeed agrees with that in \cite{PRVdS}.

%
%

\section{Conclusions and Discussions}\label{sec:conclusion}

In this paper we have calculated the correlators of the stress-energy tensor, or the noise kernel in the theory of stochastic gravity, in $R^{N}$ and $AdS^{N}$ spacetimes. The method we have used is the generalized zeta-function regularization procedure devised in \cite{PH97}. The zeta-function regularization method was originally used to deal mainly with one loop effective action and the corresponding operators associated with it, notably the expectation value of the stress-energy tensor. The method introduced in \cite{PH97} enables one to deal also with operator correlation functions. To understand the correlations more closely and also to facilitate further applications, we have developed their small and large geodesic distance limits. Particularly in the small geodesic distance limit, we could verify explicitly that the correlators satisfy the conservation equation. Moreover, if the coupling is conformal, the correlators also satisfy the traceless conditions.

We have compared our results with that in \cite{OsbSho00} and \cite{PRVdS}. The agreement shows that the zeta-function method used here is an useful alternative to evaluate correlators of operators. Applications of interest to us include the physics of AdS-black holes. To do that we shall first extend our consideration to the finite temperature case, and which is the subject of our next paper.

In our calculation we have considered correlators of the stress-energy tensor of two different spacetime points. If the coincident limit is taken, we would then obtain fluctuations of the expectation values of the stress-energy tensor. In that case further divergences will develop, as can be seen from our small distance expansions. In fact, in \cite{PH97} fluctuations of the energy density of a thermal scalar field was considered. There examples were chosen cleverly so that no further regularization was needed to arrive at a finite answer. However, in general one needs to deal with these divergences with further regularization, in much the same manner as one tries to generalize the zeta-function regularization to higher-loop situations. This has already been achieved by the so-called operator regularization initiated by McKeon and Sherry \cite{McKShe87}.

Even when operator regularization is implemented, there is still another complication concerning contact terms. In the conservation equation the covariant derivative on the correlator of the stress-energy tensor is non-zero but proportional to contact terms involving delta-functions \cite{OsbSho00}. For regularization methods like the dimensional regularization, delta-functions are regularized to zero and the regularized covariant derviative indeed goes to zero. However, for the zeta-function regularization, or the operator regularization, delta-functions are not regularized to zero. One must therefore enforce conservation by hand. As discussed also in \cite{OsbSho00} the conservation of the correlator in maximally symmetric spaces enables one to express the correlator itself in terms of two scalar functions, corresponding to spin-0 and spin-2 contributions. Regularization imposing on these two functions can be done in such a way that conservation is manifest in the whole procedure. We hope to come back to this in more detail in the future.

%
%

\acknowledgments

HTC is supported in part by the National Science Council of the Republic of China under the Grant NSC 99-2112-M-032-003-MY3, and by the National Center for Theoretical Sciences. BLH is supported in part by the National Science Foundation under grants PHY-0601550 and PHY-0801368 to the University of Maryland.



\appendix

%
%

\section{Small geodesic distance expansions for various dimensions in AdS}\label{appendix}

In this Appendix we list the small geodesic distance expansions of the coefficients $C_{i}(N)$ in the correlators of the stress-energy tensors in $AdS^{N}$ for $N=2$ to $11$, except $N=4$ and $5$ which are already presented in Section \ref{sec:limits}.

First we list the results for even dimensions. For $N=2$,
\begin{eqnarray}
C_{1}(2)&=&\frac{1}{\pi^{2}\sigma^{4}}\left[2(1-12\xi+22\xi^{2})
-48\xi^{2}\left(\gamma+\psi\left(\frac{1}{2}+a\sqrt{b}\right)
+{\rm ln}\left(\frac{\sigma}{2a}\right)\right)\right]\ \ \ \ \ \ \ \ \ \ \ \ \ \ \ \ \nonumber\\
&&+\frac{1}{12\pi^{2}a^{2}\sigma^{2}}
\bigg\{(-1+18\xi+2\xi^{2})+24\xi(1-\xi)\left(\gamma+\psi\left(\frac{1}{2}+a\sqrt{b}\right)+{\rm ln}\left(\frac{\sigma}{2a}\right)\right)\nonumber\\
&&\ \ \ \ \ \ \ \ +12a^{2}b\left[(-1+10\xi-6\xi^{2})-8\xi(1-\xi)\left(\gamma+\psi\left(\frac{1}{2}+a\sqrt{b}\right)+{\rm ln}\left(\frac{\sigma}{2a}\right)\right)\right]\bigg\}\nonumber\\
&&+\cdots\nonumber\\
C_{2}(2)&=&\frac{1}{\pi^{2}\sigma^{4}}\left[2\xi(2-7\xi)
+8\xi^{2}\left(\gamma+\psi\left(\frac{1}{2}+a\sqrt{b}\right)+{\rm ln}\left(\frac{\sigma}{2a}\right)\right)\right]\nonumber\\
&&+\frac{1}{48\pi^{2}a^{2}\sigma^{2}}\bigg\{(-9+16\xi-28\xi^{2})\nonumber\\
&&\ \ \ \ \ \ \ \ \ \ \ +2(-3+12\xi-44\xi^{2})\left(\gamma+\psi\left(\frac{1}{2}+a\sqrt{b}\right)+{\rm ln}\left(\frac{\sigma}{2a}\right)\right)\nonumber\\
&&\ \ \ \ \ \ \ \ \ \ \ +12a^{2}b\left[(-1+4\xi^{2})+2(1-2\xi)^{2}\left(\gamma+\psi\left(\frac{1}{2}+a\sqrt{b}\right)+{\rm ln}\left(\frac{\sigma}{2a}\right)\right)\right]\bigg\}\nonumber\\
&&+\cdots\nonumber\\
C_{3}(2)&=&\frac{1}{2\pi^{2}\sigma^{4}}\left[(1-8\xi+12\xi^{2})-16\xi^{2}\left(\gamma+\psi\left(\frac{1}{2}+a\sqrt{b}\right)+{\rm ln}\left(\frac{\sigma}{2a}\right)\right)\right]\nonumber\\
&&+\frac{1}{96\pi^{2}a^{2}\sigma^{2}}\bigg\{2(-7+32\xi-28\xi^{2})\nonumber\\
&&\ \ \ \ \ \ \ \ \ \ \ +2(-3+24\xi-8\xi^{2})\left(\gamma+\psi\left(\frac{1}{2}+a\sqrt{b}\right)+{\rm ln}\left(\frac{\sigma}{2a}\right)\right)\nonumber\\
&&\ \ \ \ +24a^{2}b\left[(-1+8\xi-4\xi^{2})+(1-8\xi+8\xi^{2})\left(\gamma+\psi\left(\frac{1}{2}+a\sqrt{b}\right)+{\rm ln}\left(\frac{\sigma}{2a}\right)\right)\right]\bigg\}\nonumber\\
&&+\cdots\nonumber\\
C_{4}(2)&=&\frac{1}{4\pi^{2}\sigma^{4}}\left[(1-2\xi)^{2}-8\xi^{2}\left(\gamma+\psi\left(\frac{1}{2}+a\sqrt{b}\right)+{\rm ln}\left(\frac{\sigma}{2a}\right)\right)\right]\nonumber\\
&&+\frac{1}{96\pi^{2}a^{2}\sigma^{2}}\bigg\{(-13+52\xi-48\xi^{2})\nonumber\\
&&\ \ \ \ \ \ \ \ \ \ \ +2(-3+12\xi+4\xi^{2})\left(\gamma+\psi\left(\frac{1}{2}+a\sqrt{b}\right)+{\rm ln}\left(\frac{\sigma}{2a}\right)\right)\nonumber\\
&&\ \ \ \ +12a^{2}b\left[(-1+4\xi)+2(1-2\xi)^{2}\left(\gamma+\psi\left(\frac{1}{2}+a\sqrt{b}\right)+{\rm ln}\left(\frac{\sigma}{2a}\right)\right)\right]\bigg\}\nonumber\\
&&+\cdots\nonumber\\
C_{5}(2)&=&-\frac{1}{4\pi^{2}\sigma^{4}}\left[(1+4\xi-36\xi^{2})+8\xi^{2}\left(\gamma+\psi\left(\frac{1}{2}+a\sqrt{b}\right)+{\rm ln}\left(\frac{\sigma}{2a}\right)\right)\right]\nonumber\\
&&+\frac{1}{96\pi^{2}a^{2}\sigma^{2}}\bigg\{(19-44\xi-80\xi^{2})\nonumber\\
&&\ \ \ \ \ \ \ \ \ \ \ +4(3-18\xi+50\xi^{2})\left(\gamma+\psi\left(\frac{1}{2}+a\sqrt{b}\right)+{\rm ln}\left(\frac{\sigma}{2a}\right)\right)\nonumber\\
&&\ \ \ \ +12a^{2}b\left[3(1-4\xi)-4(1-6\xi+6\xi^{2})\left(\gamma+\psi\left(\frac{1}{2}+a\sqrt{b}\right)+{\rm ln}\left(\frac{\sigma}{2a}\right)\right)\right]\bigg\}\nonumber\\
&&+\cdots
\end{eqnarray}

For $N=6$,
\begin{eqnarray}
C_{1}(6)&=&\frac{24}{\pi^{6}\sigma^{12}}(3-28\xi+70\xi^{2})-\frac{3}{\pi^{6}a^{2}\sigma^{10}}
\left[(3-22\xi+50\xi^{2})+4a^{2}b(1-10\xi+30\xi^{2})\right]+\cdots\nonumber\\
C_{2}(6)&=&-\frac{24}{\pi^{6}\sigma^{12}}(1-5\xi)^{2}+\frac{1}{8\pi^{6}a^{2}\sigma^{10}}
\left[(59-580\xi+1420\xi^{2})+20a^{2}b(1-6\xi)^{2}\right]+\cdots\nonumber\\
C_{3}(6)&=&\frac{6}{\pi^{6}\sigma^{12}}(1-8\xi+20\xi^{2})\nonumber\\
&&-\frac{1}{16\pi^{6}a^{2}\sigma^{10}}
\left[(27-184\xi+440\xi^{2})+20a^{2}b(1-8\xi+24\xi^{2})\right]+\cdots\nonumber\\
C_{4}(6)&=&\frac{1}{\pi^{6}\sigma^{12}}(1-4\xi+10\xi^{2})-\frac{1}{16\pi^{6}a^{2}\sigma^{10}}
\left[(7-28\xi+68\xi^{2})+4a^{2}b(1-4\xi+12\xi^{2})\right]+\cdots\nonumber\\
C_{5}(6)&=&\frac{1}{\pi^{6}\sigma^{12}}(13-132\xi+330\xi^{2})\nonumber\\
&&-\frac{1}{4\pi^{6}a^{2}\sigma^{10}}
\left[(29-291\xi+721\xi^{2})+4a^{2}b(1-13\xi+39\xi^{2})\right]+\cdots
\end{eqnarray}

For $N=8$,
\begin{eqnarray}
C_{1}(8)&=&\frac{1152}{\pi^{8}\sigma^{16}}(1-9\xi+21\xi^{2})\nonumber\\
&&-\frac{12}{\pi^{8}a^{2}\sigma^{14}}
\left[(25-212\xi+490\xi^{2})+4a^{2}b(3-28\xi+70\xi^{2})\right]+\cdots\nonumber\\
C_{2}(8)&=&-\frac{48}{\pi^{8}\sigma^{16}}(3-14\xi)^{2}+\frac{1}{2\pi^{8}a^{2}\sigma^{14}}
\left[(393-3654\xi+8491\xi^{2})+84a^{2}b(1-5\xi)^{2}\right]+\cdots\nonumber\\
C_{3}(8)&=&\frac{24}{\pi^{8}\sigma^{16}}(3-24\xi+56\xi^{2})\nonumber\\
&&-\frac{1}{8\pi^{8}a^{2}\sigma^{14}}
\left[(255-1896\xi+4396\xi^2)+84a^{2}b(1-8\xi+20\xi^{2})\right]+\cdots\nonumber\\
C_{4}(8)&=&\frac{3}{\pi^{8}\sigma^{16}}(3-12\xi+28\xi^{2})\nonumber\\
&&-\frac{1}{8\pi^{8}a^{2}\sigma^{14}}
\left[(45-180\xi+418\xi^{2})+12a^{2}b(1-4\xi+10\xi^{2})\right]+\cdots\nonumber\\
C_{5}(8)&=&\frac{6}{\pi^{8}\sigma^{16}}(39-366\xi+854\xi^{2})\nonumber\\
&&-\frac{1}{8\pi^{8}a^{2}\sigma^{14}}
\left[(1317-12324\xi+28690\xi^{2})+12a^{2}b(13-132\xi+330\xi^{2})\right]+\cdots
\end{eqnarray}

For $N=10$,
\begin{eqnarray}
C_{1}(10)&=&\frac{5760}{\pi^{10}\sigma^{20}}(5-44\xi+99\xi^{2})\nonumber\\
&&-\frac{240}{\pi^{10}a^{2}\sigma^{18}}
\left[(49-421\xi+945\xi^{2})+12a^{2}b(1-9\xi+21\xi^{2})\right]+\cdots\ \ \ \ \ \ \ \ \ \ \nonumber\\
C_{2}(10)&=&-\frac{2880}{\pi^{10}\sigma^{20}}(2-9\xi)^{2}\nonumber\\
&&+\frac{3}{\pi^{10}a^{2}\sigma^{18}}
\left[(2351-21132\xi+47484\xi^{2})+36a^{2}b(3-14\xi)^{2}\right]+\cdots\nonumber
\end{eqnarray}
\begin{eqnarray}
C_{3}(10)&=&\frac{1440}{\pi^{10}\sigma^{20}}(1-8\xi+18\xi^{2})\nonumber\\
&&-\frac{3}{2\pi^{10}a^{2}\sigma^{18}}
\left[(581-4456\xi+10008\xi^{2})+36a^{2}b(3-24\xi+56\xi^{2})\right]+\cdots\nonumber\\
C_{4}(10)&=&\frac{144}{\pi^{10}\sigma^{20}}(1-4\xi+9\xi^{2})\nonumber\\
&&-\frac{3}{2\pi^{10}a^{2}\sigma^{18}}
\left[(77-308\xi+692\xi^{2})+4a^{2}b(3-12\xi+28\xi^{2})\right]+\cdots\nonumber\\
C_{5}(10)&=&\frac{144}{\pi^{10}\sigma^{20}}(43-388\xi+873\xi^{2})\nonumber\\
&&-\frac{3}{\pi^{10}a^{2}\sigma^{18}}
\left[(1777-16018\xi+36010\xi^{2})+4a^{2}b(39-366\xi+854\xi^{2})\right]+\cdots\nonumber\\
\end{eqnarray}

Next, we present the results for odd dimensions. For $N=3$,
\begin{eqnarray}
C_{1}(3)&=&\frac{3}{8\pi^{2}\sigma^{6}}\left(3-32\xi+128\xi^{2}\right)
-\frac{3}{8\pi^{2}a^{2}\sigma^{4}}\left[4\xi+a^{2}b^{2}\left(1-16\xi\right)\right]+\cdots\nonumber\\
C_{2}(3)&=&-\frac{3}{16\pi^{2}\sigma^{6}}(1-8\xi)^{2}+\frac{1}{16\pi^{2}a^{2}\sigma^{4}}\left[(1-8\xi+32\xi^{2})-2a^{2}b\right]+\cdots\nonumber\\
C_{3}(3)&=&\frac{3}{16\pi^{2}\sigma^{6}}(1-8\xi+32\xi^{2})-\frac{1}{8\pi^{2}a^{2}\sigma^{4}}
\left[\xi(1+4\xi)+a^{2}b(1-8\xi)\right]+\cdots\nonumber\\
C_{4}(3)&=&\frac{1}{16\pi^{2}\sigma^{6}}(1-4\xi+16\xi^{2})-\frac{1}{16\pi^{2}a^{2}\sigma^{4}}
\left[4\xi^{2}+a^{2}b(1-4\xi)\right]+\cdots\nonumber\\
C_{5}(3)&=&\frac{1}{16\pi^{2}\sigma^{6}}(1-24\xi+96\xi^{2})-\frac{1}{16\pi^{2}a^{2}\sigma^{4}}
\left[(1-12\xi+40\xi^{2})-2a^{2}b(1-4\xi)+\cdots\right]\nonumber\\
\end{eqnarray}

For $N=7$,
\begin{eqnarray}
C_{1}(7)&=&\frac{315}{128 \pi^6 \sigma^{14}} (35 - 320 \xi + 768 \xi^2)\nonumber\\
&&\ \ \ \ \ -\frac{105}{128 \pi^6 a^2  \sigma^{12}}
 \left[4 (5 - 41 \xi + 96 \xi^2) + 3 a^2 b (5 - 48 \xi + 128 \xi^2)\right]+\cdots\nonumber\\
C_{2}(7)&=&-\frac{315}{256 \pi^6 \sigma^{14}} (5 - 24 \xi)^2\nonumber\\
&&\ \ \ \ \ +\frac{15}{256 \pi^{6} a^2 \sigma^{12}}
 \left[(199  - 1896 \xi + 4512 \xi^2)+ 6 a^2 b (3 - 16 \xi)^2\right]+\cdots\nonumber\\
 C_{3}(7)&=&\frac{315}{256 \pi^6 \sigma^{14}} (5 - 40 \xi + 96 \xi^2)\nonumber\\
 &&\ \ \ \ \ -\frac{15}{128 \pi^{6} a^2 \sigma^{12}}
 \left[ (19 - 137 \xi + 324 \xi^2) + 3 a^2 b (3 - 24 \xi + 64 \xi^2)\right]+\cdots\nonumber\\
C_{4}(7)&=&\frac{45}{256 \pi^6 \sigma^{14}} (5 - 20 \xi + 48 \xi^2)\nonumber\\
&&\ \ \ \ \ -\frac{15}{256 \pi^{6}a^2 \sigma^{12}}
\left[4(2 - 8 \xi + 19 \xi^2) + a^2 b (3 - 12 \xi + 32 \xi^2)\right]+\cdots\nonumber
\end{eqnarray}
\begin{eqnarray}
C_{5}(7)&=&\frac{45}{256 \pi^6 \sigma^{14}} (95 - 920 \xi + 2208 \xi^2)\nonumber\\
&&\ \ \ \ \ -\frac{15}{256\pi^{6} a^2 \sigma^{12}}
 \left[(179 - 1724 \xi + 4120 \xi^2) + 8 a^2 b (3 - 33 \xi + 88 \xi^2)\right]+\cdots\nonumber\\
 \end{eqnarray}

For $N=9$,
\begin{eqnarray}
C_{1}(9)&=&\frac{14175}{512 \pi^8 \sigma^{18}} (63 - 560 \xi + 1280 \xi^2)\nonumber\\
&&\ \ \ \ \ -\frac{2835}{512 \pi^{8} a^2 \sigma^{16}}
 \left[(105 - 900 \xi + 2048 \xi^2 )+ a^2 b (35 - 320 \xi + 768 \xi^2)\right]+\cdots\nonumber\\
C_{2}(9)&=&-\frac{14175}{1024 \pi^8 \sigma^{18}}(7 - 32 \xi)^2\nonumber\\
&&\ \ \ \ \ +\frac{315}{1024\pi^{8} a^2 \sigma^{16}}\left[(1175 - 10720 \xi + 24448 \xi^2)+ 8 a^2 b (5 - 24 \xi)^2 \right]+\cdots\nonumber\\
C_{3}(9)&=&\frac{14175}{1024 \pi^8 \sigma^{18}} (7 - 56 \xi + 128 \xi^2)\nonumber\\
&&\ \ \ \ \ -\frac{315}{1024 \pi^{8}a^2 \sigma^{16}}\left[(165 - 1250 \xi + 2848 \xi^2) + 8 a^2 b (5 - 40 \xi + 96 \xi^2)\right]+\cdots\nonumber\\
C_{4}(9)&=&\frac{1575}{1024 \pi^8 \sigma^{18}} (7 - 28 \xi + 64 \xi^2)\nonumber\\
&&\ \ \ \ \ -\frac{315}{1024 \pi^{8} a^2 \sigma^{16}}\left[ (25 - 100 \xi + 228 \xi^2) + a^2 b (5 - 20 \xi + 48 \xi^2)\right]+\cdots\nonumber\\
C_{5}(9)&=&\frac{1575}{512 \pi^8 \sigma^{18}}(119 - 1092 \xi + 2496 \xi^2)\nonumber\\
&&\ \ \ \ \ -\frac{315}{1024 \pi^{8}a^2\sigma^{16}}\left[ (930 - 8520 \xi + 19448 \xi^2 )+
    a^2 b (95 - 920 \xi + 2208 \xi^2)\right]+\cdots\nonumber\\
\end{eqnarray}

For $N=11$,
\begin{eqnarray}
C_{1}(11)&=&\frac{3274425}{2048 \pi^{10} \sigma^{22}} (33 - 288 \xi + 640 \xi^2)\nonumber\\
&&\ \ \ \ \ -\frac{155925}{2048\pi^{10} a^2 \sigma^{20}}\left[4 (84 - 721 \xi + 1600 \xi^2)+a^2 b (63 - 560 \xi + 1280 \xi^2)\right]+\cdots
\nonumber\\
C_{2}(11)&=&-\frac{1091475}{4096 \pi^{10} \sigma^{22}} (9 - 40 \xi)^2\nonumber\\
&&\ \ \ \ \ +\frac{4725}{4096\pi^{10} a^2 \sigma^{20}}\left[ (12957  - 115080 \xi + 255520 \xi^2)+ 30 a^2 b (7 - 32 \xi)^2\right]+\cdots\nonumber\\
C_{3}(11)&=&\frac{1091475}{4096 \pi^{10} \sigma^{22}} (9 - 72 \xi + 160 \xi^2)\nonumber\\
&&\ \ \ \ \ -\frac{4725}{2048\pi^{10}a^2 \sigma^{20}}\left[ (714 - 5523 \xi + 12260 \xi^2) +
    15 a^2 b (7 - 56 \xi + 128 \xi^2)\right]+\cdots\nonumber\\
C_{4}(11)&=&\frac{99225}{4096 \pi^{10} \sigma^{22}} (9 - 36 \xi + 80 \xi^2)\nonumber\\
&&\ \ \ \ \ -\frac{4725}{4096 \pi^{10}a^2 \sigma^{20}}\left[4 (42 - 168 \xi + 373 \xi^2)+ 3 a^2 b (7 - 28 \xi + 64 \xi^2)\right]+\cdots\nonumber
\end{eqnarray}
\begin{eqnarray}
C_{5}(11)&=&\frac{99225}{4096 \pi^{10} \sigma^{22}} (477 - 4248 \xi + 9440 \xi^2\nonumber\\
&&-\frac{4725}{4096 a^2 \pi^{10} \sigma^{20}}\left[ (9429 - 83916 \xi + 186376 \xi^2) +
    6 a^2 b (119 - 1092 \xi + 2496 \xi^2)\right]\nonumber\\
    &&\ \ \ +\cdots\nonumber\\
\end{eqnarray}

%
%


\end{document}